\documentclass[11pt]{article}
\usepackage{amsmath,amssymb,color,graphics,epsfig,cite}
\usepackage{graphicx,subfigure}

\usepackage{amsfonts}

\newcommand{\hoch}[1]{$\, ^{#1}$}

\newcommand{\be}{\begin{equation}}
	\newcommand{\ee}{\end{equation}}
\newcommand{\bea}{\setlength\arraycolsep{2pt} \begin{eqnarray}}
	\newcommand{\eea}{\end{eqnarray}}
\newcommand{\nn}{\nonumber}

\def\ft#1#2{{\textstyle{\frac{\scriptstyle #1}{\scriptstyle #2} } }}
\def\fft#1#2{{\frac{#1}{#2}}}

\def\0{{\sst{(0)}}}
\def\1{{\sst{(1)}}}
\def\2{{\sst{(2)}}}
\def\3{{\sst{(3)}}}
\def\4{{\sst{(4)}}}
\def\5{{\sst{(5)}}}
\def\6{{\sst{(6)}}}
\def\7{{\sst{(7)}}}
\def\8{{\sst{(8)}}}
\def\sst#1{{\scriptscriptstyle #1}}

\begin{document}
	
\begin{center}

{\Large {\bf Kasner Singularity of Black Holes in Einstein-scalar Gravity}}
		
\vspace{20pt}
		
Ze-Xuan Xiong\hoch{1} and H. L\"u\hoch{2,1}
		
\vspace{10pt}

{\it \hoch{1}The International Joint Institute of Tianjin University, Fuzhou,\\ Tianjin University, Tianjin 300350, China}

\medskip

{\it \hoch{2}Center for Joint Quantum Studies, Department of Physics,\\
School of Science, Tianjin University, Tianjin 300350, China }

\vspace{40pt}
		
\underline{ABSTRACT}
\end{center}

We study the spacelike Kasner singularity of spherically-symmetric, static and asymptotically flat black holes in Einstein gravity minimally coupled to a massless scalar with a suitable self-interacting scalar potential. We focus on how the asymptotic information such as the mass and scalar charge affect the properties of the Kasner singularity, including the Kasner exponents. We show how a nontrivial integration constant can be extracted from the near-singularity geometry and find a general pattern that this integration constant asymptotes to a linear combination of the mass and scalar charge at large mass limit. We also find that there may be a black hole upper bound on the maximum surviving time of a massive particle inside such a black hole before it falls into the Kasner singularity, and the Schwarzschild black hole saturate this bound.

\vfill {\footnotesize xzx\_167@tju.edu.cn\ \ \ mrhonglu@gmail.com}
	
	\thispagestyle{empty}
	\pagebreak

	\tableofcontents
	\addtocontents{toc}{\protect\setcounter{tocdepth}{2}}
	
	\newpage

\section{Introduction}
\label{sec:intro}

Black hole singularities represent one of the most puzzling and intriguing aspects of General Relativity. It has become a fundamental issue after Penrose proved the singularity theorem under the strong energy condition (SEC) \cite{Hawking:1973uf,Senovilla:2014gza}, which states that the black hole geodesic is incomplete, but the nature of the incompleteness is not generally clear. For the Schwarzschild and Reissner-Nordstr\"om (RN) black holes, the singularity is spacelike and timelike respectively, and it appears as the divergence of the curvature tensor. There can also be impenetrable boundaries in the form of irrational power in the near-horizon geometry of some rotating extremal black holes \cite{Mao:2023qxq,Mao:2025nrp}, which resembles a whimper in cosmology \cite{Ellis:1974ug,siklos}. The cosmic censorship forbids the existence of the timelike singularity, but it remains a strong cosmic censorship conjecture (SCCC) of Penrose \cite{sccc1,sccc2,sccc3}.

While the exterior physics of black holes is well understood, their interior structure remains far less explored, and one typically appeals the resolution of the singularity to quantum gravity. However, the large domain of a black hole interior still belongs to the classical region, as the majority of time, as much as close to $\pi M$, of a massive particle motion inside a Schwarzschild black hole of mass $M$, can be described geodesically before it falls into the singularity. It is thus instructive to study the black hole interior classically despite of the singularity. The SCCC of Penrose motivates us to focus our attention on black holes with spacelike singularity, and the Schwarzschild black hole is the simplest and best known such example \cite{Regge:1957td,Hawking:1970zqf, Ruffini:1971bza}. Indeed, the interior of the Schwarzschild black hole exhibits cosmological anisotropy and terminates at a spacelike singularity \cite{Kasner:1921zz,Belinsky:1970ew, Belinskii:1973sud,Henneaux:2007ej}. The singularity core naturally leads to the need for a specialized solution, such as the Kasner metric \cite{Kasner:1921zz}, which describes the nature of cosmological singularities and has since been applied to study the spacelike singularity of black holes without a Cauchy horizon, e.g.~\cite{Hartnoll:2020fhc,Liu:2021hap,Henneaux:2022ijt,An:2022lvo, Auzzi:2022bfd,Liu:2022rsy,Gao:2023zbd, Cai:2023igv,Albrychiewicz:2024gti,Zhang:2025tsa, Zhang:2025hkb}.

   Taking the Schwarzschild black hole as an example, the Kasner-type singularity core is characterized by the universal exponents that are independent of the black hole mass. An important question is whether such a Kasner structure is a special feature of Schwarzschild
black holes, or instead represents a more universal description of black hole singularities. This question becomes particularly compelling when matter fields are introduced. Although in string or string inspired theories, many black hole solutions involving the dilatonic scalar continue to have both the inner and outer horizons, introducing a scalar matter field generally destroys the inner horizon, or the horizon altogether in the extremal limit \cite{Isenberg:2015rqa,Ringstrom:2015jza,Geng:2018jck,
Hartnoll:2020rwq,Cai:2020wrp,Grandi:2021ajl,An:2021plu}. This causes the black hole interiors to become solely cosmological and to terminate at a spacelike singularity, just as in the Schwarzschild case, although the Kasner structures are now modified by the scalar or other matter fields.

  It is thus natural to investigate whether the Kasner structure persists in black hole spacetimes coupled to scalar fields. Indeed, recent studies have begun to uncover rich and diverse interior structures when scalars are involved, as illustrated in the literature listed so far. However, most existing works focus on charged black holes with charged scalar hair in increasingly complicated theories. The interior structure of scalar hairy black hole solutions supported solely by a real scalar has received comparatively little attention. We believe this is worth a study on its own, not simply because the Einstein-scalar theory is relatively simple, involving only one real scalar. It is also nontrivial since it involves the subtleties of no hairy theorem \cite{Bekenstein:1995un}. A plethora of evidences indicate that black holes can carry a real scalar hair, see, e.g.~a review \cite{Herdeiro:2015waa}, but the scalar hair parameter is secondary in that it is not a continuous independent parameter, but a function of mass or other conserved charges, satisfying the weaker version of the no-scalar hair theorem conjecture \cite{Lu:2025eub}.

This weaker version of the no-scalar hair theorem conjecture provides a strong restriction on the black hole interior structure, since the number of parameters describing the interior geometry of the scalar hairy black hole must also be reduced. Specifically, as we study the black hole interior from the event horizon to the singularity core, we have to first make sure that the horizon belongs to a black hole, which can be smoothly connected to asymptotic infinity. The weeding out of the non-black hole interior is not only paramount to present a correct picture of black hole interior, but also allows us to study how the exterior properties manifest themselves in the interior and affect its near-singularity geometry. For this reason, we shall focus in this paper on the study of asymptotically flat black holes, instead of asymptotically anti-de Sitter (AdS) black holes. One reason is that the definition of black holes that are asymptotic to AdS spacetimes is much looser in the literature, as many asymptotically locally AdS geometries with an event horizon are also referred to as AdS black holes, as we can see in the listed references earlier. The other reason is that the asymptotically flat black holes are more relevant to astrophysical black holes that may be observed.

In this work, we aim to investigate the Kasner singularity of spherically-symmetric and static black holes in Einstein gravity minimally coupled to a real massless scalar $\phi$,
focusing on how the scalar affects the interior geometry and the relationship between asymptotic and singularity parameters. For the theory to admit a scalar hairy black hole, there must be a suitable concave potential to evade the standard (stronger version of) no-hair theorem. The general spherically-symmetric and static solution contains two independent integration constants or parameters. For asymptotically flat geometries, they manifest themselves as the mass $M$ and scalar hair parameter $\Sigma$, entering the solution asymptotically as
\be
-g_{tt} \sim \fft{1}{g_{rr}} \sim 1 - \fft{2M}{r} + \cdots \,,\qquad
\phi \sim \fft{2\Sigma}{r} + \cdots\,.\label{twop}
\ee
For lexical simplicity, we shall refer to $\Sigma$ as a ``scalar charge'', a common usage in literature, even though the scalar is not a gauge field. For the solution to describe a black hole, $\Sigma$ is not an independent parameter, but a specific function of the mass $M$. Once the mass is given, the geometry of the black hole, both outer and interior regions, is completely determined. This raises two questions
\begin{itemize}

\item Will the Kasner exponents at the singularity depend on the asymptotic black hole parameters $(M, \Sigma(M))$?

\item How can we read off or extract these parameters near the singularity, as they can be easily read off at the asymptotic region?

\end{itemize}
In order to answer these two questions, focusing only on the possible Kasner structures that can arise in the Einstein-scalar theory is not enough. This is because, as in the case of the asymptotic behavior, the solution in the local region contains spurious non-black hole parameters. We need to first construct the scalar hairy black hole solutions in the theory. Although Einstein-scalar gravity is one of the simplest theories, there is a great freedom of constructing a scalar potential that can give rise to a black hole solution, whose geometry depends on the scalar potential. Thus, the classification of scalar hairy black holes can be a formidable task in Einstein-scalar gravity. Furthermore, exact solutions are rare, but numerical solutions are cumbersome to construct as they all involve a fine-tuning of the scalar charge $\Sigma(M)$. We therefore consider both exact analytical solutions in literature \cite{Anabalon:2012ta,Anabalon:2013qua,Gonzalez:2013aca, Acena:2013jya,Feng:2013tza}, and general considerations that apply to broad classes of Einstein-scalar theories by constructing and testing some new specific numerical examples. We find that a general pattern of the Kasner singularity appears to emerge.

The paper is organized as follows. In Section~\ref{sec:ES}, we introduce the general framework of Einstein-scalar gravity, present the equations of motion for spherically-symmetric and static black holes. We show that there are generally no inner horizons. We derive the general structure of Kasner-type singularities for the interior geometry. We also define the key quantities, including the Kasner exponents and the near-singularity parameter $\Xi$, that will be used throughout the paper. In Section \ref{sec:exact}, we review a class of exact scalar hairy black hole solutions in literature. We examine their exterior structures and the thermodynamic properties. We then study the interior structures and determine the Kasner exponents, in terms of the parameters of the theory, and establish a linear relation between $\Xi$ and $(M, \Sigma)$ that characterizes the asymptotic falloffs. We also demonstrate the existence of a mass gap for kinetic-dominated solutions. In Section~\ref{sec:general}, we extend our analysis to a general class of scalar potentials, and set up a numerical approach for constructing both the black hole exterior and its interior. The basic approach is to first construct the near-horizon geometry, which involves two free nontrivial parameters. We integrate the horizon geometry to asymptotic infinity, which requires a fine-tuning of these two parameters. Once the black hole horizon is determined, we can integrate to both outwardly to the asymptotic infinity to read off the mass and scalar charge, and inwardly to the singularity to read off the near-singularity properties. In Section \ref{sec:numerical}, we summarize our numerical findings of our new numerical black holes. In particular, we find that there exist multiple branches of solutions where the scalar charge $\Sigma$ is a discrete function of the mass. We compare our numerical findings to those we obtained from the exact solutions, which allows us to present some general patterns. In Section \ref{sec:interiorgeo}, we probe the interior geometry by studying the geodesic motion of a massive particle and determine the maximum proper time that a particle can survive before falling into the Kasner singularity. Finally, we conclude in Section \ref{sec:conclusion} with a discussion of our main results and future directions. In Appendix \eqref{sec:app}, we show that the Kasner singularity involving a kinetic-energy dominating scalar can be understood from Kasner singularity in higher dimensional pure gravity via the Kaluza-Klein approach.

\section{Einstein-scalar gravity}
\label{sec:ES}

\subsection{Theory and equations}

In this paper, we consider Einstein gravity minimally coupled to a massless scalar with a self-interacting potential. The Lagrangian takes the general form
\be
{\cal L}=\sqrt{-g} \big(R - \ft12 (\partial\phi)^2 - V(\phi)\big)\,.
\ee
We assume that the potential has a fixed point $\phi=0$ with the constraint
\be
V(0)=V'(0)=V''(0)=0\,.\label{fixpointcons}
\ee
This ensures that the vacuum is the Minkowskian spacetime, with $\phi=0$. Furthermore, the scalar is massless, with the leading asymptotic falloff
\be
\phi \sim \fft{2\Sigma}{r}\,,
\ee
where the integration constant $\Sigma$ is the scalar hair parameter, which we also refer to as the scalar charge for lexical simplicity, even though $\phi$ is not a gauge field.

It is clear that the theory admits the usual Schwarzschild black hole as its solution. In this paper, we study spherically symmetric and static black holes that also carry the nonzero scalar charge $\Sigma$. The general ansatz takes the form
\be
ds^2 = - h(r) dt^2 + \fft{dr^2}{f(r)} + r^2 d\Omega_2^2\,,\qquad \phi=\phi(r)\,.
\label{genmetric1}
\ee
The complete set of equations of motion for the functions $(h,f,\phi)$ is given by
\bea
&&\fft{f'}{f} + \fft{1}{r} + \fft{r^2 V-2}{2rf} + \fft14 r \phi'^2=0\,,\qquad
\fft{h'}{h} + \fft{1}{r} + \fft{r^2 V-2}{2rf} - \fft14 r \phi'^2=0\,,\nn\\
&&\phi'' + \frac{2 (f+1)- r^2V}{2 r f } \phi ' - \fft{V'(\phi)}{f}=0\,.
\eea
Here a prime denotes a derivative with respect to the radius $r$. The first two equations above also imply
\be
\phi'^2 = \fft{2}{r} \Big(\log\fft{h}{f}\Big)'\,.\label{phipsq}
\ee
Note that the scalar equation $\Box\phi = dV/d\phi$ can be cast into
\be
-r^2 \sqrt{hf} \phi'^2 - r^2\phi \sqrt{\fft{h}{f}} \fft{d V}{d\phi} =
\fft{d}{dr} (r^2 \sqrt{hf} \phi \phi')\,.
\ee
This implies that the existence of a black hole, which has a horizon at $r=r_+$ and the asymptotic falloffs $h\sim f = 1 - \fft{2M}{r} + \cdots$, requires that the scalar potential is concave, i.e.~$\phi V'(\phi) <0$. The equation \eqref{phipsq} also implies the absence of the inner horizon, since it would otherwise require
\be
\int_{r_-}^{r_+} \Big(-r^2 \sqrt{hf} \phi'^2 - r^2\phi \fft{\partial V}{\partial\phi} \sqrt{\fft{h}{f}}\Big) dr =0\,.
\ee
This is a fine-tuning condition and not satisfied in general. Thus, scalar hairy black holes in general have only one horizon, and the singularity is spacelike.

\subsection{Kasner singularity}

The absence of an inner horizon implies that the interior of a scalar hairy black hole is generally cosmological with a Kasner type singularity, namely
\be
ds^2 = -d\tau^2 + a_1^2\, \tau^{2P_t} dt^2 + a_2^2\, \tau^{2P_T} d\Omega_{2}^2\,,\qquad
\phi =2 P_\phi \log\tau\,,\label{D4kasnersol}
\ee
where the Kasner exponents $(P_t,P_T,P_\phi)$ are constants, depending on the parameters of the theory, as well as possibly integration constants of the black hole solution. (Note that ``T'' in $P_T$ stands for the ``transverse'' directions.) In the case when the scalar kinetic term dominates over its potential at the singularity, we have
\be
P_t + 2 P_T =1\,,\qquad P_\phi^2 + 3P_T^2 - 2P_T=0\,.\label{Psconstraints}
\ee
The second equation can also be written as a quadratic constraint $2P_\phi^2 + P_t^2 +2 P_T^2=1$ \cite{Kasner:1921zz,Belinskii:1973sud,Henneaux:2022ijt}. In Appendix \ref{sec:app}, we show that the two relations have a higher dimensional Kasner geometrical origin of pure gravity in the Kaluza-Klein approach. Note that the Kasner-type solution \eqref{D4kasnersol} with \eqref{Psconstraints} is only the leading-order approximate solution of Einstein gravity coupled to a free massless scalar in the $\tau\rightarrow 0$ limit; it becomes an exact solution for all $\tau$ if we replace the 2-sphere metric $d\Omega_2^2$ by a Ricci-flat metric.

As we shall see later, the relations in \eqref{Psconstraints} collapse when the scalar potential is not inferior at the singularity. We shall largely focus on the former case in this paper, but also give explicit examples where the potential energy cannot be ignored even at the singularity. To read off the Kasner exponents from the original metric and scalar functions $(h,f,\phi)$, we first define
\be
c_1 = \lim_{r\rightarrow 0} r\, (\log f)'\,,\qquad
c_2 = \lim_{r\rightarrow 0} r\, (\log h)'\,,\qquad
c_3 = \lim_{r\rightarrow 0} r \phi'\,,\label{csdef}
\ee
then the Kasner exponents are given by
\be
P_t=\fft{c_2}{2-c_1}\,,\qquad P_T=\fft{2}{2-c_1}\,,\qquad P_\phi =-\fft{c_3}{2-c_1}\,.\label{ccons}
\ee
Thus, the relations \eqref{Psconstraints} become
\be
c_1+c_2 + 2 =0\,,\qquad c_3^2=4 (c_2+1)\ge 0\,,\qquad\rightarrow\qquad c_1 + 1\le 0\,.
\ee
This implies that the Kasner exponent $P_T$ is always positive and bounded above
\be
P_T \le \fft23\,.\label{PTbound}
\ee
For the Schwarzschild black hole, we have $c_1=-1=c_2$ and $c_3=0$, and hence the Kasner exponents for the Schwarzschild black hole are $P_t=-1/3$, $P_T=2/3$ and $P_\phi=0$. Thus
we see that the Schwarzschild black hole saturate the bound \eqref{PTbound}. For the general case, we substitute $(c_1,c_2)$ of \eqref{csdef} into the first equation above, and integrate it out, we have
\be
\Xi \equiv \lim_{r\rightarrow 0} r \sqrt{h f} = \hbox{finite constant.}\label{xidef}
\ee
It is easy to verify, using \eqref{phipsq}, that the second equation of \eqref{ccons} gives no further information. It is clear that constant parameter $\Xi$ is the manifestation of the black hole asymptotic hair at the singularity. For example, the Schwarzschild black hole has only one parameter, namely the mass $M$, and we have
\be
\Xi=2M\,,\qquad \hbox{for the Schwarzschild black hole.}
\ee
One of our goals in this paper is to study how the asymptotic hair parameters, such as mass $M$ and scalar charge $\Sigma$, are encoded in this singularity parameter, namely $\Xi=\Xi(M,\Sigma)$.

It is worth mentioning that in many exact solutions, it is more convenient to write the solution in a different radial coordinate gauge,
\be
ds^2= - h(r) dt^2 + \fft{dr^2}{f(r)} + R(r)^2 d\Omega_2^2\,, \qquad \phi = \phi(r)\,.
\label{newmetric}
\ee
In this case, we have
\be
c_1= \lim_{r\rightarrow 0} \fft{R (R'^2 f)'}{R'^3 f}\,,\qquad
c_2 =  \lim_{r\rightarrow 0} \fft{R h'}{R' h}\,,\qquad
c_3 =\lim_{r\rightarrow 0} \fft{R}{R'}\phi'\,,\qquad
\Xi=\lim_{r\rightarrow 0} R R' \sqrt{h f}\,.
\ee

Finally, we emphasize again that the constant $\Xi$ exists only for singularities where the kinetic-energy of the scalar is dominating, with the Kasner exponents satisfying \eqref{Psconstraints}. For general Kasner-type singularities, we can always extract a dimensionless parameter, given by
\be
\Theta = \lim_{r\rightarrow 0} (-f)^{P_t}\, (-h)^{1-P_T}\,,\qquad \hbox{or}\qquad
\Theta = \lim_{r\rightarrow 0} (-R'^2 f)^{P_t}\, (-h)^{1-P_T}\,.\label{Thetadef}
\ee
for the metrics \eqref{genmetric1} or \eqref{newmetric} respectively. For the Schwarzschild black hole, we have $\Theta_{\rm sch} =1$. However, we do not find this parameter as enlightening as the $\Xi$ parameter for general scalar hairy black holes.

\section{Exact scalar hairy black holes}
\label{sec:exact}

\subsection{Scalar potential and exact black hole solutions}

In this section, we study the properties of the Kasner singularity of a class of exact scalar hairy black holes, constructed by the reversing engineer technique \cite{Anabalon:2012ta, Anabalon:2013qua,Gonzalez:2013aca,Acena:2013jya,Feng:2013tza}. We follow the notation of \cite{Feng:2013tza} and the scalar potential is given by
\bea
V &=& - \fft{\alpha}{\mu(1-4\mu^2)}
\bigg(2 \sinh \left(\frac{\mu  \phi }{\nu }\right) \left(\left(2 \mu ^2+1\right) \cosh \left(\frac{\phi }{\nu }\right)-2 \mu ^2+2\right)\nn\\
&&\qquad\qquad\qquad-6 \mu  \sinh \left(\frac{\phi }{\nu }\right) \cosh \left(\frac{\mu  \phi }{\nu }\right)\bigg)\,,\label{scalarpot}
\eea
where $\alpha$ is the coupling constant, and $\nu=\sqrt{1-\mu^2}$ is a dimensionless parameter. This gives rise to a one-parameter ($\mu$) family of Einstein-scalar theories, which may provide some universal structure of the Kasner singularity of Einstein-scalar theories. Note that the sign choice of $\nu$ can be absorbed into the coupling parameter $\alpha$. The potential vanishes at $\phi=0$. The first few leading-order terms of the small $\phi$ Taylor expansion are
\be
V=-\fft{\alpha}{30 \left(1-\mu ^2\right)^{3/2} }\left(\phi ^5+\frac{\left(3 \mu ^2+2\right) }{42(1- \mu ^2)} \phi ^7+\frac{\left(2 \mu ^4+4 \mu ^2+1\right)}{1008 \left(\mu ^2-1\right)^2} \phi ^9 +O\left(\phi ^{10}\right)\right).\label{Vtaylor}
\ee
Thus, the potential satisfies the fixed point condition \eqref{fixpointcons}. Note that there exist smooth limits of $\mu=0, \pm 1/2$ \cite{Feng:2013tza}.  In particular, the $\mu=0$ limit of the scalar potential was first proposed in \cite{Zloshchastiev:2004ny}.

The Einstein-scalar theory with the scalar potential \eqref{scalarpot} admits a general class of exact scalar hairy black hole solutions in the form of \eqref{newmetric}, with \cite{Feng:2013tza}
\bea
h= f &=& H^{-\mu} \bigg(1 - \fft{ \alpha r^2}{2\mu(1-4\mu^2)}
\Big(-H^{2 \mu +1}+ \mu  (2 \mu +1) H^2-4  \mu ^2 H +H+\mu  (2 \mu -1)\Big)\bigg)\,,\nn\\
&& \phi = \sqrt{1-\mu^2}\log H\,,\qquad R=r H^{\fft12(1+\mu)}\,,\qquad
H=1 + \fft{q}{r}\,.\label{exactgen}
\eea
Note that the solution is invariant under
\be
q\rightarrow -q\,,\qquad \mu\rightarrow -\mu\,,\qquad r \rightarrow r-q\,.
\ee
We can thus choose, without loss of generality that $q\ge 0$ and $\mu \in [-1,1)$. Note that $\mu=\pm 1$ both reduce to the Schwarzschild black hole; however, $\mu\rightarrow -1$ is a smooth limit, but $\mu\rightarrow 1$ is not. We thus exclude the $\mu=1$ from the consideration.

\subsection{Exterior properties}

The general solution \eqref{exactgen} contains only one integration constant $q$. The metric is asymptotic to Minkowski spacetime at the infinity. We can read off the mass $M$ and scalar charge $\Sigma$ from the asymptotic falloff behaviors. We find
\be
M=\ft1{12} \alpha q^3 + \ft12\mu q\,,\qquad \Sigma=\ft12 \sqrt{1-\mu^2} q\,.
\ee
Thus, we see that $q$ is a natural parameter for the scalar charge. The solution describes a black hole when the integration constant $q$ satisfies the condition
\be
q > q_{\rm min}\,,\qquad
q_{\rm min}=
\left\{
  \begin{array}{ll}
    \sqrt{\fft{2(1-2\mu)}{\alpha}}\,, & \qquad -1 \le \mu<\ft12\,, \\
    0\,, &\qquad\ \  \ft12 \le \mu<1\,.
  \end{array}
\right.\label{qmin}
\ee
When this condition is satisfied, there exists one and only one positive root $r_0$ for the metric function $h=f$, i.e.~$h(r_0)=0=f(r_0)$. We can use the standard technique to determine the temperature and entropy, given by
\be
T=\fft{f'(r_0)}{4\pi}\,,\qquad S=\pi r_0^2 H(r_0)^{1+\mu}\,,
\ee
It is then straightforward to verify that the first law of black hole thermodynamics holds, namely
\be
dM=T dS\,.\label{firstlaw}
\ee
It is clear that the black hole \eqref{exactgen} is not the most general spherically symmetric and static solutions, since the scalar charge $\Sigma$  is not an independent parameter, but a function of mass.

The expressions for the mass and temperature are functions of the entropy, as indicated by the first law; however, they are complicated in expressions, and there are no analytic forms. In the large $q\rightarrow \infty$ limit, the black hole approaches the Schwarzschild black hole, with the mass/entropy relation
\be
q=\frac{\sqrt[3]{6r_0}}{\sqrt[3]{\alpha }}+
\frac{1-\mu }{\alpha ^{2/3} \sqrt[3]{6r_0}} - \frac{1-\mu ^2}{10 \alpha  r_0} + \cdots\,.
\ee
\be
M=\frac{\sqrt{S}}{2 \sqrt{\pi }} + \frac{3\ 3^{2/3} \sqrt[6]{\pi } \left(1-\mu ^2\right)}{40 \sqrt[3]{2} \alpha ^{2/3} \sqrt[6]{S}} -\frac{\sqrt[3]{3} \pi ^{5/6} \left(1-\mu ^2\right) \left(31 \mu ^2+9\right)}{2240\ 2^{2/3} \alpha ^{4/3} S^{5/6}} + \cdots\,.
\ee
Thus, we see that the general solution, at large mass limit, approaches the Schwarzschild black hole. Indeed, the scalar charge/mass ratio approaches zero as the mass $M$ increases to infinity.

The opposite small mass limit, $r_0\to 0$, is more complicated, and the $\mu$-dependence of the solution is sensitive.  As $r_0$ is dialled down toward zero, the scalar charge $q$ decreases, but may refuse to vanish, as we see in \eqref{qmin}. In this $r_0\rightarrow 0$ regime, the radius $r_0$ shrinks in lock-step with $q$, and the asymptotic relation between the two is set by the theory parameter $\mu$:
\bea
-1\le \mu<0: &&q=\frac{\sqrt{2(1 - 2\mu)}}{\sqrt{\alpha}} -\fft{r_0}{2\mu} + \cdots\,, \nn\\
0 \le \mu<\ft12: &&q=\frac{\sqrt{2(1 - 2\mu)}}{\sqrt{\alpha}} + \frac{2^{\mu-1} \left(1-2\mu\right)^{\mu} r_0^{1-2 \mu}}{\alpha^{\mu}\mu (1+2 \mu)} \cdots\,, \nn\\
\ft12\le \mu< 1 :&&q=2^{\frac{1}{2 \mu+1}} \mu^{\frac{1}{2 \mu+1}} \left(4 \mu^2 -1\right)^{\frac{1}{2 \mu+1}} \alpha^{-\frac{1}{2 \mu+1}} r_0^{\frac{2 \mu-1}{2 \mu+1}} + \cdots\,.
\eea
Expanding $M$ in powers of $S=\pi R(r_0)^2$, we find three distinct behaviours, each controlled by parameter $\mu$:
\bea
-1\le\mu<0:&&
M=\frac{(1+\mu)\sqrt{1-2\mu}}{3\sqrt{2\alpha}}+
\frac{(\mu-1)\pi^{1/(\mu-1)}}{4\mu}\Bigl(
\frac{\alpha}{2-4\mu}\Bigr)^{\!-\frac{1+\mu}{2(\mu-1)}}S^{\frac{1}{1-\mu}}
+\cdots\,,\nn\\
[4pt] 0\le\mu<\tfrac12:&&
M=\frac{(1+\mu)\sqrt{1-2\mu}}{3\sqrt{2\alpha}}-
\frac{(\mu-1)(2-4\mu)^{\frac{1-3\mu}{2(\mu-1)}}
\pi^{\frac{1-2\mu}{\mu-1}}}{4\mu(2\mu+1)\alpha^{\frac{3\mu-1}{2(\mu-1)}}}
S^{\frac{1-2\mu}{1-\mu}}+\cdots\,,\nn\\
[4pt]\tfrac12\le\mu<1:&&
M=2^{\frac12(\frac1\mu-3)}\mu^{\frac{1+\mu}{2\mu}}(4\mu^2-1)^{
\frac12(\frac1\mu-1)}\pi^{\frac1{2\mu}-1}\alpha^{\frac{\mu-1}{2\mu}}
S^{1-\frac1{2\mu}}+\cdots\,.
\eea
Note that for $-1\le \mu<\ft12$, the leading term is independent of $S$. In other words, as the black hole approaches zero size with $S=0$, the mass does not go to zero, in contrast to the Schwarzschild black hole. Thus, these scalar hair black holes have a mass gap
\be
M_{\text{gap}}=\frac{(1+\mu)\sqrt{1-2\mu}}{3\sqrt{2\alpha}}\,, \qquad \mu\in(-1,\tfrac12)\,.
\ee
While the Schwarzschild black holes exist for all mass, the spectrum of the scalar hairy black holes is gapped, and they exist only for $M>M_{\rm gap}$. On the other hand, for $\mu\ge 1/2$, the mass slides continuously to zero as $S\to 0$.  This sharp transition at $\mu=1/2$  marks the boundary between a ``massive floor'' and a ``massless'' phase in the parameter space of the theories.

\subsection{Interior properties}

The interior of the black holes becomes cosmological and the evolution ends in a class of Kasner singularities. We find that the Kasner exponents are solely determined by the parameter $\mu$, namely
\bea
-1\le \mu<\ft12: &&\qquad P_t = \fft{\mu}{2-\mu}\,,\qquad
P_T = \fft{1-\mu}{2-\mu}\,,\qquad P_\phi = \fft{\sqrt{1-\mu^2}}{2-\mu}\,,\nn\\
\ft12\le \mu< 1: &&\qquad  P_t = \fft{1-\mu}{1+\mu}=P_T\,,\qquad
P_3 = \sqrt{\fft{1-\mu}{1+\mu}}\,.\label{kasnerexp}
\eea
Note that in all the cases, $P_T$ is positive and bounded above as in \eqref{PTbound}.
For $\mu\in [-1,1/2)$, the scalar kinetic term dominates over the scalar potential and the Kasner exponents satisfy the relations given in \eqref{Psconstraints}. We can verify this fact by computing the ratio $ V(\phi)/(\partial\phi)^2 $ near the Kasner singularity, given by
\be
 \lim_{r\to 0} \fft{V(\phi)}{(\partial\phi)^2 } =
 \left\{
\begin{array}{ll}
 \lim_{r\to 0} \frac{\alpha  q r}{\mu(\mu -1)   \left(4 \mu +\alpha  q^2-2\right)} = 0, & \qquad -1\le \mu<0\,,\\
& \\
\lim_{r \to 0} \fft{4\alpha q r \log(r)}{2- \alpha q^2} = 0, &\qquad \mu=0\,,\\
&\\
\lim_{r \to 0} \frac{\alpha  (1-2 \mu ) q^{2 \mu +1} r^{1-2 \mu }}{\mu  (\mu +1) (2 \mu +1) \left(4 \mu +\alpha  q^2-2\right)} = 0, &\qquad 0< \mu<\ft12\,.
\end{array}
 \right.
\ee
On the other hand, for $\mu\in (1/2,1)$,  the limit tends to a finite and nonzero value
\be
  \lim_{r\to 0} \fft{V(\phi)}{(\partial\phi)^2 } = \fft{2\mu-1}{1 + \mu}\, .
\ee
Hence $\mu = 1/2$  not only serves as the critical point between scalar hairy black holes with and without mass gap, but also swaps the Kasner exponents by precisely demarcates the domain of kinetic-dominated theory, as indicated in \eqref{kasnerexp}. It is worth remarking that based on the ansatz \eqref{D4kasnersol}, the scalar kinetic energy necessarily dominates at the singularity if the scalar potential is of the polynomial type;
however, the scalar potential discussed in this section involves exponentials such that the potential may not always be ignored at the singularity, as we see for the case with $\mu > 1/2$.

For the kinetic-dominated class of solutions $(-1\le \mu<\ft12)$, we have finite $\Xi$, given by
\be
\Xi = \fft{(1-\mu)}{4(1-2\mu)} \hat \alpha q^3 - \ft12 (1-\mu) q=
\fft{3(1-\mu)}{1-2\mu} M - \fft{\sqrt{1-\mu^2}}{1-2\mu} \Sigma\,.
\ee
This suggests the possible  existence of a universal linear relation between the singularity parameter $\Xi$ and the asymptotic parameters, the mass and scalar charge, namely
\be
\Xi=\xi_1 M + \xi_2 \Sigma\,,\label{XiMS}
\ee
where $(\xi_1, \xi_2)$ are constants depending only on the theory, but independent of the integration constants of a solution. Furthermore, the scalar hairy black holes satisfy an inequality
\be
\xi_1 = \fft{\partial \Xi}{\partial M} = \fft{3(1-\mu)}{1-2\mu}\ge 2\,.
\ee
The bound is saturated by the Schwarzschild black hole.

Note that for the general $\mu$, we can extract a dimensionless parameter $\Theta$ given by
\eqref{Thetadef}. For the exact solutions, we find
\bea
\Theta =
\left\{
  \begin{array}{ll}
    2^{\frac{2 \mu }{\mu -2}} (1-\mu )^{-\frac{2 \mu }{\mu -2}} \left(\frac{\alpha  q^2}{2 (1-2 \mu )}-1\right)^{\frac{1+\mu}{2-\mu }}, & \qquad -1 \le \mu<\ft12\,, \\
&\\
    \frac{2^{\frac{\mu -3}{\mu +1}} (1-\mu )^{\frac{2 (1-\mu)}{1+\mu}}}{\mu  \left(4 \mu ^2-1\right)} \alpha q^2, & \quad\qquad \ft12 <\mu <1\,.
  \end{array}
\right.
\eea
The constant $\Theta$ is divergent at $\Theta =\fft12$. To understand this dimensionless parameter, it is worth note that for the exact solutions, we have
\be
\fft16 \alpha q^2 = \sqrt{1-\mu^2} \fft{M}{\Sigma} - \mu\,.
\ee
Thus, the exact solutions illustrate that $\Theta$ is a function of asymptotic dimensionless ratio $M/\Sigma$. While it is of interest to extract the nontrivial parameter $\Theta$ from the Kasner singularity, its dependence on $M/\Sigma$ is too complicated to be enlightening.

\section{General cases}
\label{sec:general}

In the previous section, we study exact black hole solutions in a class of Einstein-scalar theories with one continuous parameter $\mu$. All these black hole interior has the Kasner singularity. The Kasner exponents all depend only on the theory parameter $\mu$, but independent of the mass, an integration constant of the solutions. In the case where the scalar kinetic term dominates over its potential, we have a linear relation \eqref{XiMS}. A natural question arises: Are these properties universal? Exact solutions are very special, and may not be representative of the more general scalar hairy black holes. Furthermore, for a given mass, there is only one scalar hairy black hole, while the weak no-scalar hair conjecture states that the scalar hair parameter cannot be an independent continuous parameter, it can still be a discrete function of mass $M$. However, we find no evidence of additional scalar hairy black holes in the previous section.

In this section, we shall address these issues by constructing further new numerical black hole examples.

\subsection{Asymptotic behavior and scalar potential}

For simplicity, we begin with a scalar potential of the polynomial type $V=-g_n \phi^n$ where $g_n$ is a coupling constant. To satisfy the condition \eqref{fixpointcons}, we must have $n\ge 3$. Furthermore, we would like to have a large-$r$ asymptotic behaviors in the form of integer-powered falloffs
\bea
f &=& 1 + \frac{\hat{f}_1}{r} + \frac{\hat{f}_2}{r^2} + \frac{\hat{f}_3}{r^3} + \cdots\,,\nn \\
h &=& 1 + \frac{\hat{h}_1}{r} + \frac{\hat{h}_2}{r^2} + \frac{\hat{h}_3}{r^3} + \cdots\,,\nn \\
\phi &=& \frac{\hat{\phi}_1}{r} + \frac{\hat{\phi}_2}{r^2} + \frac{\hat{\phi}_3}{r^3} + \cdots\,,\label{larger}
\eea
where the mass and scalar charge are given by $M = -\hat{h}_1/2, ~ \Sigma= \hat{\phi}_1 /2 $. Substituting this ansatz into the equations of motion and solving them order by order in the large-$r$ expansion, we find that we must have $n\ge 5$.

  It turns out from our numerical analysis that a single-term potential will not lead to a scalar hairy black hole. Inspired by \eqref{Vtaylor} from the scalar potentials with exact black hole solutions, we consider
\be
V=-g_5 \phi^5 - g_7 \phi^7\,.\label{g57}
\ee
The potential can be viewed as the small $\phi$ approximation of the one considered in the previous section. With \eqref{g57}, the equations of motion can be solved order by order in the large-$r$ power-series expansion. We find that all the coefficients can be solved in terms of the mass $M$ and scalar charge $\Sigma$. Specifically, we find the leading falloffs are given by
\be
\begin{aligned}
\hat{h}_1 & = \hat{f}_1\,,\qquad \hat{f}_2 = \frac{1}{4}\hat{\phi}_1^2\,,\qquad \hat{h}_2 = 0\,,\qquad \hat{\phi}_2 = -\frac{1}{2}\hat{\phi}_1(\hat{f}_1 + 5g_5\hat{\phi}_1^3)\,, \\
\hat{f}_3 &= -\frac{1}{8}\hat{\phi}_1^2(\hat{f}_1 + 12g_5\hat{\phi}_1^3)\,,\qquad \hat{h}_3 = -\frac{1}{24}(\hat{f}_1\hat{\phi}_1^2 + 4g_5\hat{\phi}_1^5)\,, \\
\hat{\phi}_3 &= \frac{1}{24}(8\hat{f}_1^2 - \hat{\phi}_1^2 + 80g_5\hat{f}_1\hat{\phi}_1^3 + 200g_5^2\hat{\phi}_1^6)\,.\label{largercoeff}
\end{aligned}
\ee

\subsection{Near-horizon geometry}

We are interested in black hole solutions with an event horizon, located at $r_0$. We assume that the near-horizon geometry is analytic, with
\bea
f(r) &=& f_1 (r-r_0) +  f_2 (r-r_0)^2 + \cdots\,, \nn \\
h(r) &=&  (r-r_0) +  h_2 (r-r_0)^2 + \cdots\,, \nn \\
\phi(r) &=& \phi_0 +  \phi_1 (r-r_0) +  \phi_2 (r-r_0)^2 + \cdots\,.
\eea
The coefficients of the Taylor series expansion can be solved order by order in terms of the power of $(r-r_0)$. For the leading coefficients, we have
\bea
f_1 &=& \frac{1}{2} r_0 \phi _0^5 \left(g_7 \phi _0^2+g_5\right)+\frac{1}{r_0}\,,\qquad
\phi_1 = -\frac{2 r_0 \phi _0^4 \left(7 g_7 \phi _0^2+5 g_5\right)}{g_7 r_0^2 \phi _0^7+g_5 r_0^2 \phi _0^5+2}\,,\nn\\
f_2 &=& -\frac{3 r_0^4 \phi _0^8 \left(7 g_7 \phi _0^2+5 g_5\right){}^2+4 r_0^2 \phi _0^5 \left(g_7 \phi _0^2+g_5\right)+8}{4 r_0^2 \left(r_0^2 \phi _0^5 \left(g_7 \phi _0^2+g_5\right)+2\right)}\,,\nn\\
h_2 &=& \frac{r_0^4 \phi _0^8 \left(7 g_7 \phi _0^2+5 g_5\right){}^2-4 r_0^2 \phi _0^5 \left(g_7 \phi _0^2+g_5\right)-8}{2 r_0 \left(r_0^2 \phi _0^5 \left(g_7 \phi _0^2+g_5\right)+2\right){}^2}\,,\nn\\
\phi_2 &=& \fft{r_0^2 \phi _0^7 \left(7 g_7 \phi _0^2+5 g_5\right)}{\left(r_0^2 \phi _0^5 \left(g_7 \phi _0^2+g_5\right)+2\right){}^3}\Big(2 g_7 \left(\phi_0^2+42\right) \phi_0^2+2 g_5 \left(\phi_0^2+20\right)\nn\\
&&+ r_0^2 \phi _0^5 \left(g_7^2 \left(\phi _0^2-7\right) \phi _0^4+2 g_5 g_7 \left(\phi _0^2-4\right) \phi _0^2 +g_5^2 \left(\phi_0^2-5\right)\right)\Big)\,.
\eea
We can in principle obtain all the coefficients in terms of horizon radius $r_0$ and scalar hair $\phi_0$ on the horizon.

At first sight, we have two free parameters on the horizon $(r_0,\phi_0)$ and two free parameters in the asymptotic region, namely $(M,\Sigma)$. Naively, one might expect that these lead to black holes with two independent parameters, since the number of free parameters of the near-horizon geometry matches precisely that in the asymptotic region. However, numerical analysis indicates that the near-horizon geometry of a generic pair of the parameters $(r_0,\phi_0)$ cannot be integrated out to infinity, but terminates at certain spacetime singularity at some finite $r$, as illustrated in the bottom left and right panels in Fig.~\ref{r0=1solution}. Equivalently, for a generic choice of $(M,\Sigma)$, the asymptotic Minkowski spacetime will lead to a naked spacetime singularity.

\subsection{Numerical approach}

In order to construct a black hole, a fine-tuning process must be employed. We can adopt the shooting method where we integrate from the near-horizon geometry specified by $(r_0,\phi_0)$ to large $r$. Specifically, we can fix a parameter, say $r_0=1$, then adjust the horizon scalar hair $\phi_0$ carefully so that the solution can reach the asymptotic infinity. We employ this numerical shooting method and obtain the $\phi_0$ as a function of $r_0$. We then use the data-fitting technique to read of the $(M,\Sigma)$ as functions of $r_0$ from the asymptotic falloffs \eqref{larger}. We can test the validity of numerical solution by verifying the first two constraints in \eqref{largercoeff}. Having obtained the $(r_0,\phi_0)$ that connect the horizon to asymptotic infinity, we then integrate the solution from the horizon to the $r\rightarrow 0$ singularity. We can then read off all the Kasner exponents $(P_t,P_T,P_\phi)$ and the constant $\Xi$.

\begin{figure}[ht]
    \centering
     \includegraphics[width=0.6\linewidth]{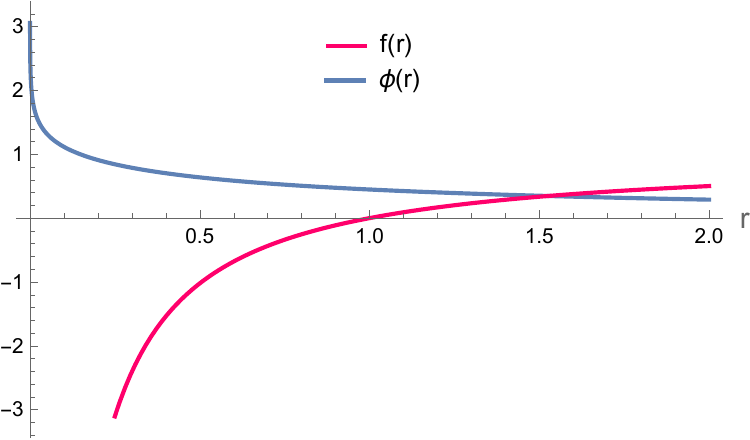}
    \includegraphics[width=0.45\linewidth]{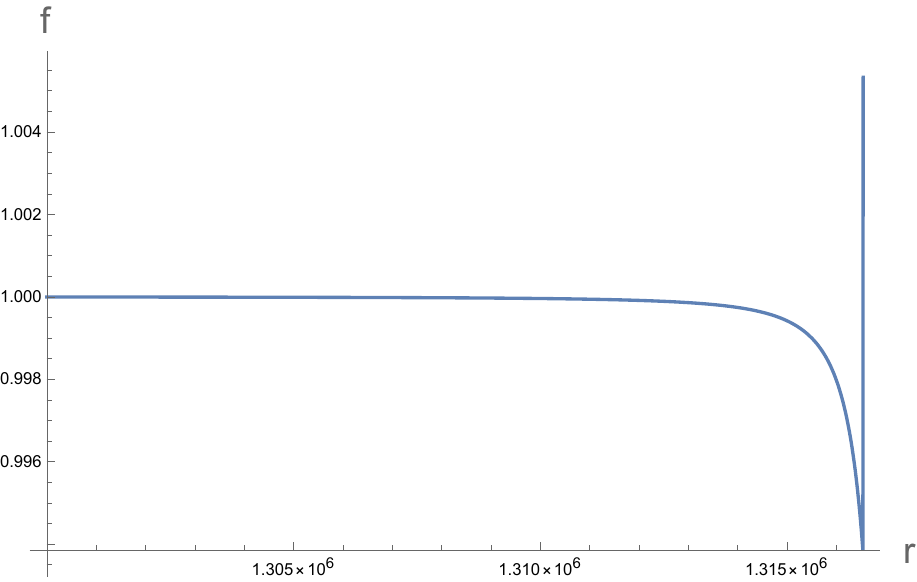}\ \ \
    \includegraphics[width=0.45\linewidth]{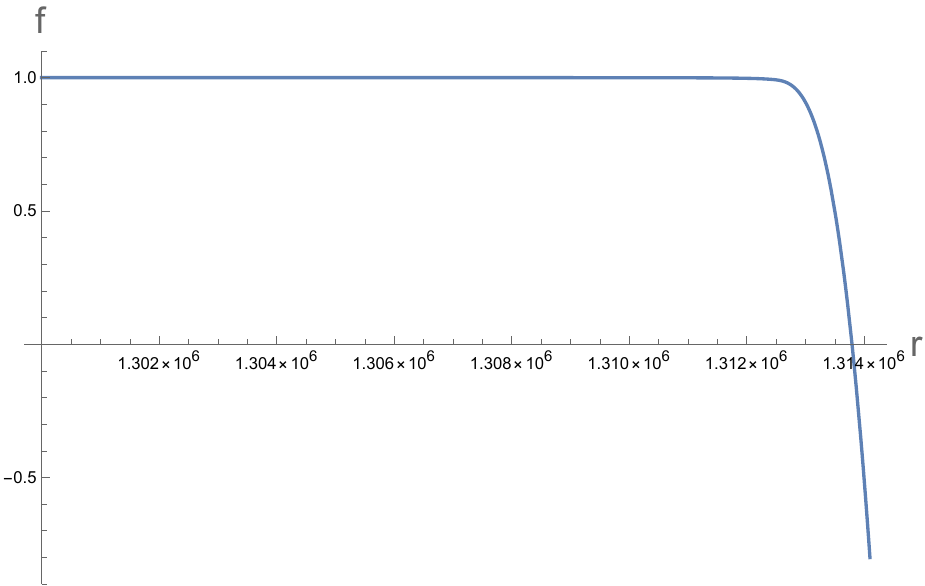}
    \caption{\small In all three graphs, we have chosen $r_0=1$. The top panel describes a black hole with $\phi_0= 0.44802$. The left and right panels describe solutions that do not have asymptotic Minkowsk region. The left panel has $\phi_0=0.50000$ and the function $f$ diverges positively at some finite $r$, whilst the right panel has $\phi_0=0.44686$ and the function $f$ diverges negatively at some finite $r$.}
    \label{r0=1solution}
\end{figure}

We now describe a concrete example of the numerical shooting method. In this example, we fix the coupling constants $g_5=1=g_7$ and choose $r_0=1$. After a careful analysis, we find that we must have $\phi_0=0.448016$, at the six significant figures. The metric function $f$ and scalar $\phi$ are plotted in the top panel in Fig.~\ref{r0=1solution}. (The bottom left and right panels illustrate how the function diverges at some large but finite $r$ when $\phi_0$ deviates from the fine-tuned value that gives to a black hole.)
We do not plot the function $h$ in the top panel of Fig.~\ref{r0=1solution}, since in the depicted region, the function $h$ and $f$ almost coalesce. However, they diverge at small $r$, where the scalar field becomes divergent. From the numerical solution, we can read off the physical quantities of interest; they are
\bea
&&\{M, \Sigma, T, P_t, P_T, P_\phi, \Xi\}\nn\\
&=&\{0.512034, 0.436951, 0.0790576, -0.324795, 0.662397,
-0.0921071, 0.979771\}\,,
\eea
with the entropy simply $S=\pi/4$. It is easy to verify that the above Kasner exponents satisfy the identities of \eqref{Psconstraints} in high accuracies. In the next section, we present a summary of our numerical results.

\section{Numerical results}
\label{sec:numerical}

It turns out that even for a simple scalar potential \eqref{g57}, an immensely rich black hole spectrum emerges. For simplicity, we begin with $g_5=1=g_7$. We find that for a given $r_0$, multiple $\phi_0$'s can exist for scalar hairy black holes, indicating that although the scalar hair is not an independent continuous parameter, it can be a discrete parameter of the mass of the black hole. In this section, we present these solutions in detail.

\subsection{Branch-1 solutions}

\begin{figure}[ht]
    \centering
     \includegraphics[width=0.6\linewidth]{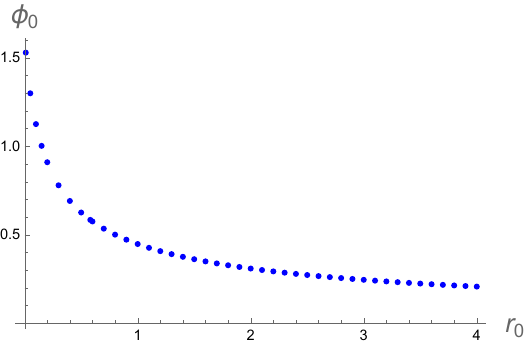}
    \includegraphics[width=0.45\linewidth]{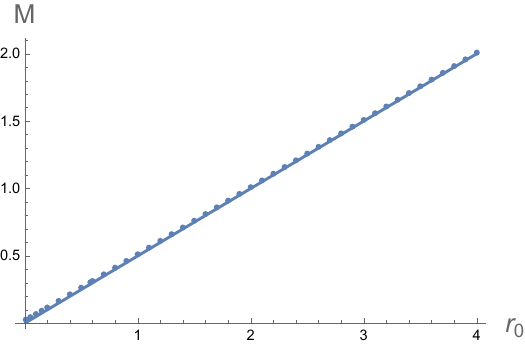}\ \ \
    \includegraphics[width=0.45\linewidth]{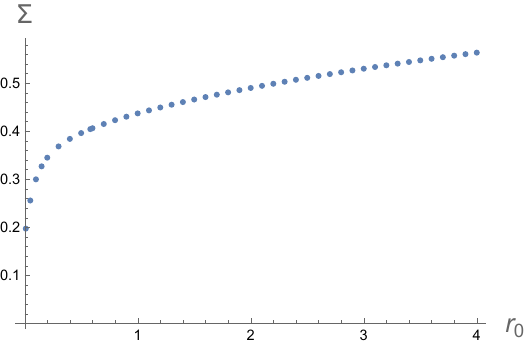}
    \caption{\small We present how the horizon scalar hair $\phi_0$, the mass $M$ and scalar charge $\Sigma$ depend on the horizon radius $r_0$. The solid line in the left panel describes the Schwarzschild black hole, which indicates that scalar hair's contribution to the spacetime geometry in the exterior region is small.}
    \label{phiMSigr0}
\end{figure}

Branch-1 black holes are those scalar hairy black holes with minimum $\phi_0$ on the horizon for a given $r_0$. The dependence of $\phi_0$, mass $M$ and the scalar charge $\Sigma$ on the horizon radius is depicted in Fig.~\ref{phiMSigr0}.
For these solutions, the scalar hair is relatively small from horizon to asymptotic infinity. It follows that the scalar contribution to the spacetime geometry is also small, such that the mass dependence of the hairy black hole is highly degenerate with the Schwarzschild black hole, as indicated in the left panel of Fig.~\ref{phiMSigr0}.

\subsubsection{Black hole thermodynamics}

\begin{figure}[ht]
    \centering
    \includegraphics[width=0.45\linewidth]{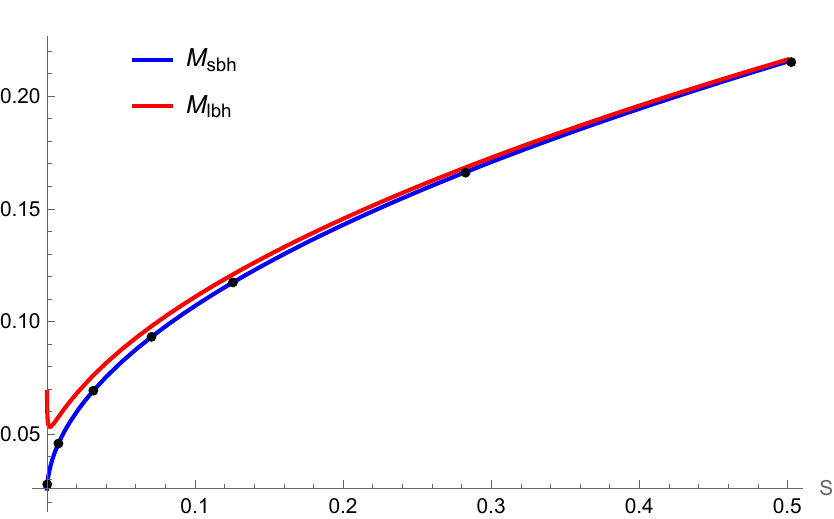}\ \ \
    \includegraphics[width=0.45\linewidth]{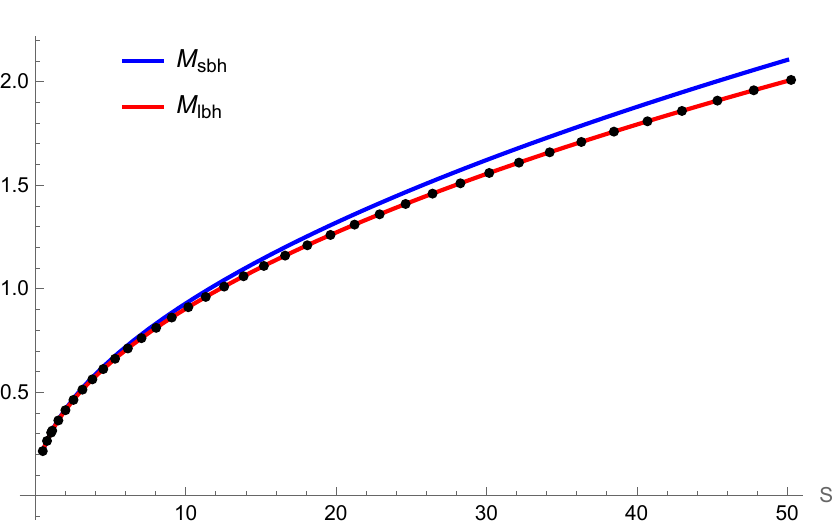}
    \caption{\small The dots are the numerical data of the mass/entropy relation. The red and blue solid lines are the mass/entropy approximate formulae \eqref{massentropy} for large and small black holes respectively. They fit the data accurately in their respective parameter regions.}
\label{MS}
\end{figure}

Since the entropy is simply given by $S=\pi r_0^2$, our numerical data allow us to determine the mass/entropy relation. For small $r_0$ or large $r_0$, we find
\bea
\hbox{small black hole}:&&\qquad M_{\rm sbh} = 0.023501 + 0.2740575 S^{0.517725}\,,\nn\\
\hbox{large black hole}:&&\qquad M_{\rm lbh} = \sqrt{\fft{S}{4\pi}} + \fft{0.0149568}{S^{0.160812}}\,.\label{massentropy}
\eea
We have constructed a large number of these new numerical black holes solutions with $r_0$ ranging from 0.01 to 4, corresponding to $S$ running from 0.0003 to around 50. We find that the mass formula for small black hole in \eqref{massentropy} fits the data well with $S\le 1$, whilst the large black hole mass formula fits the data well with $S\ge 0.5$. In Fig.~\ref{MS}, we plot the numerical data to compare with the mass/entropy formulae for both the large and small black holes. The dots are the numerical data. The blue solid line depicts the small black hole formula in \eqref{massentropy} and it indeed fits the small black hole data very well, but deviates noticeably as the entropy increases beyond $S\sim 10$. The red solid line, on the other hand, gives the large black hole formula in \eqref{massentropy}, and it fits the large black hole data accurately, but deviates from the numerical data as $S$ becomes small.

\begin{figure}[ht]
    \centering
    \includegraphics[width=0.45\linewidth]{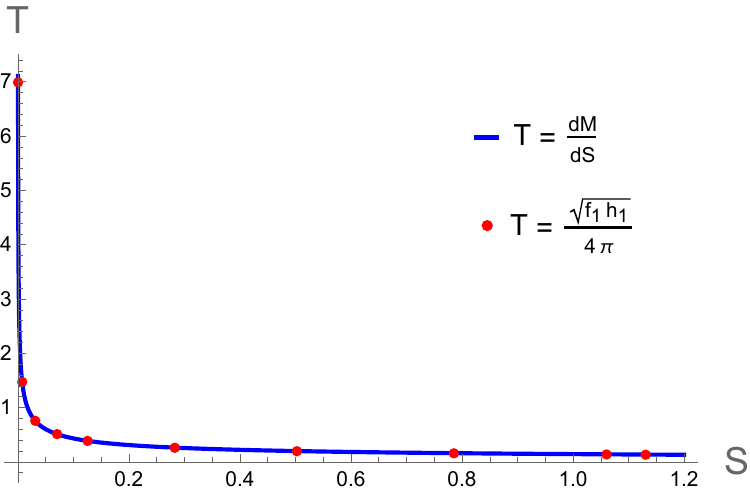}\ \ \
    \includegraphics[width=0.45\linewidth]{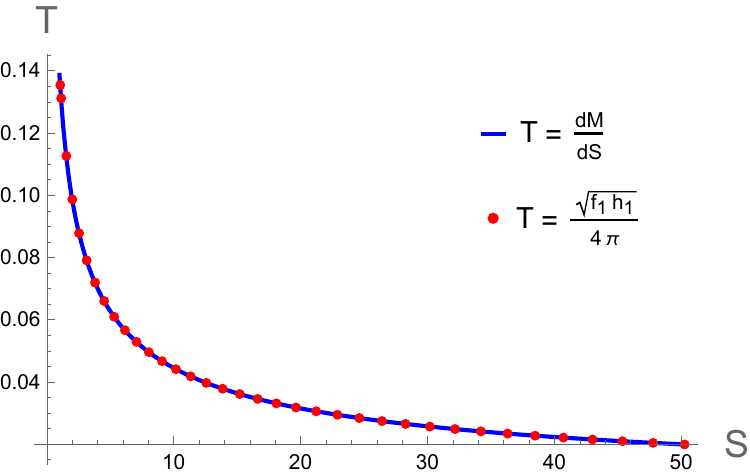}
    \caption{\small The left and right panels give the temperature as a function of entropy for both small and larger black holes respectively. The solid lines reflect the temperature based on the first law \eqref{Ttheory}, while the dots are black hole temperatures \eqref{Tdata} read off directly from the black hole numerical solutions.}
    \label{TS}
\end{figure}

Having obtained the mass/charge relation, we can test the validity of our black hole solution by checking the first law of black hole thermodynamics. We can calculate the temperature using the mass/entropy formula \eqref{massentropy} for both large and small black holes, namely
\be
T= \fft{dM}{dS}\,,\label{Ttheory}
\ee
and then compare the result with the numerical data, directly read off from the near-horizon geometry, namely
\be
T= \fft{\sqrt{h_1 f_1}}{4\pi}\,.\label{Tdata}
\ee
We present the temperature as a function of $S$ in Fig.~\ref{TS}. The solid lines are the temperature derived from \eqref{Ttheory} with the mass/entropy relations given in \eqref{massentropy}. The dots are the numerical data based on \eqref{Tdata}. We find that accuracy of the first law for all our numerical data is within 0.5\% error.

\subsubsection{Black hole interior}

\begin{figure}[ht]
    \centering
     \includegraphics[width=0.3\linewidth]{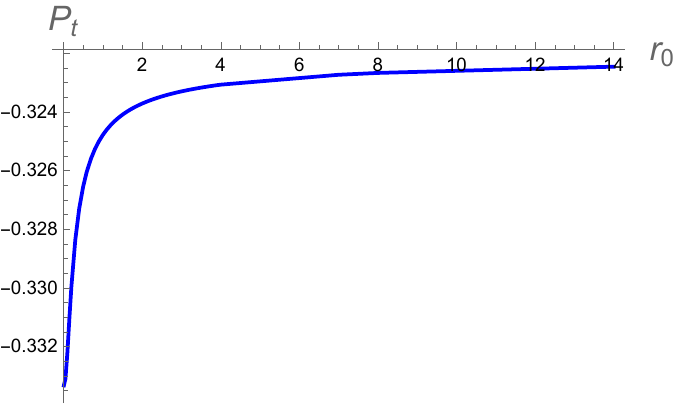}\ \ \
    \includegraphics[width=0.3\linewidth]{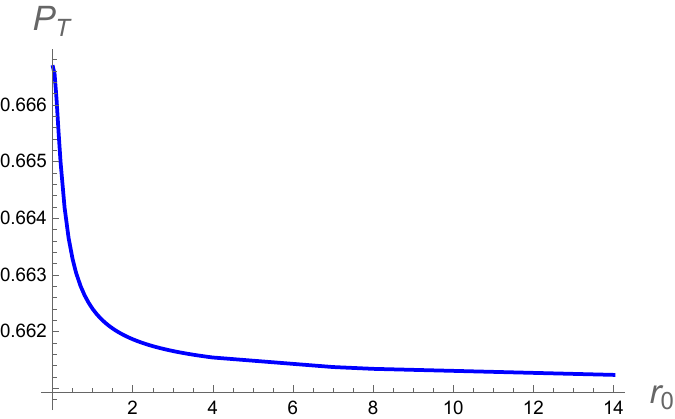}\ \ \
    \includegraphics[width=0.3\linewidth]{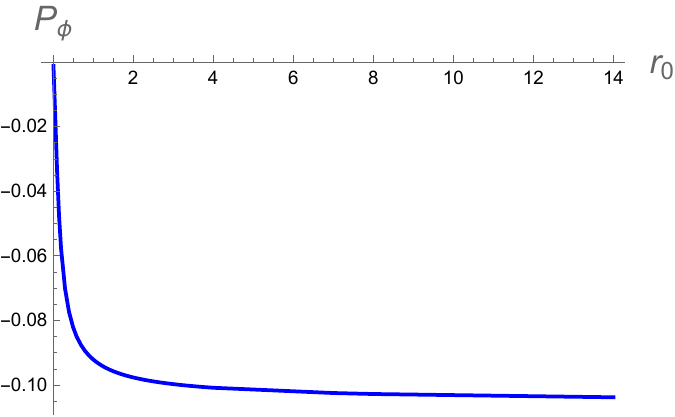}
    \caption{\small The Kasner exponents of Branch-1 solutions are functions of the black hole radius. They satisfy the two identities \eqref{Psconstraints} with extremely high accuracy ($\le 10^{-7}$.) Furthermore, $P_T$ is positive and bounded above as in \eqref{PTbound}.}
    \label{3Ps}
\end{figure}

Having verified the first law and hence established the validity of these scalar hairy black holes, we are now in the position to study their interior properties. We integrate the black hole solution from the horizon to the Kasner singularity and read off the Kasner exponents using the formulae in \eqref{csdef} and \eqref{ccons}. The data are plotted in Fig.~\ref{3Ps}. We see now that the Kasner exponents are no longer solely determined by the theory parameters, which is fixed as $g_5=1=g_7$ in these examples. Instead, the exponents depend also on  the black hole integration constant, namely the black hole horizon radius $r_0$.

As we have seen earlier, the black hole exterior geometry and the mass/entropy relation approaches those of the Schwarzschild black hole for large mass. The interior Kasner structures is resolutely different from the interior of the Schwarzschild black hole, which has $P_t=-1/3$, $P_T=2/3$ and $P_\phi=0$. Instead, the Kasner exponents resemble those of the Schwarzschild black hole when the mass is small, but they approach asymptotically some other distinct values for large black holes, as we can see in Fig.~\ref{3Ps}. It is worth remarking that in all these examples, the Kasner exponents satisfy the two identities in \eqref{Psconstraints} with an extremely high accuracy of $<10^{-7}$ in our numerical calculation. This meets the expectation, since based on the ansatz \eqref{D4kasnersol}, the scalar kinetic energy necessarily dominates over its polynomial-type of the potential energy.

To understand these specific asymptotic values of $(P_t,P_T,P_\phi)$ at large $r_0$, we note that the scalar potential \eqref{g57} can be viewed as the small $\phi$ approximation of the scalar potential \eqref{scalarpot}, as in \eqref{Vtaylor}. Thus for $g_5=1=g_7$, we have effective $\mu$, as
\be
\mu=\pm \fft{2\sqrt2}{3}\,.
\ee
When $\mu\le 1/2$, the higher-order scalar contributions become negligible even inside the black hole since the Kasner singularity is dominated by the kinetic term in this case. Thus, the Branch-1 solutions would correspond to take $\mu = -2\sqrt2/3$, if all the higher order terms could be ignored. We thus would expect that at large $r_0$, the Kasner exponents in Fig.~\ref{3Ps} should approach to
\be
\{P_t,P_T,P_\phi\}^{\mu=-2\sqrt2/3} = \{\ft17(2-3\sqrt2),\ft1{14}(5+3\sqrt2),\ft14(3-\sqrt2)\}
\sim \{-0.320,0.660,0.113\}\,.
\ee
Our numerical results are indeed close to these numbers, but with a small and noticeable deviation. We find that
\be
\lim_{r_0\rightarrow \infty} P_t - P_{t}^{\mu=-2\sqrt2/3} \sim  0.002\,.
\ee
This indicates that the black holes of the theory \eqref{g57}, when the interior is considered, differs from those of \eqref{scalarpot} (with $\mu=-2\sqrt2/3$) even at the large $r_0$ limit, but the difference is small.

\begin{figure}[ht]
    \centering
     \includegraphics[width=0.45\linewidth]{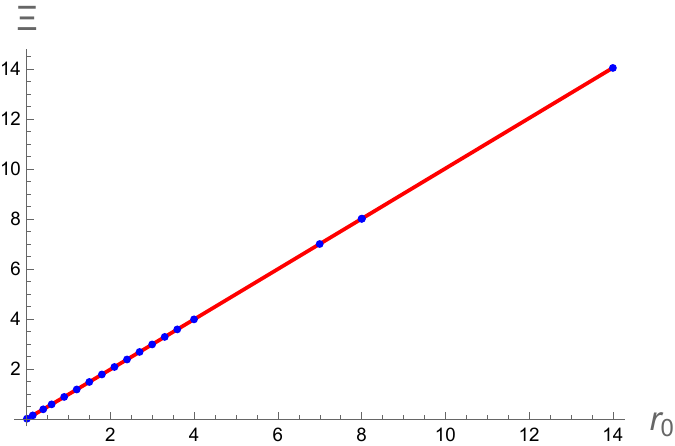}\ \ \
\includegraphics[width=0.45\linewidth]{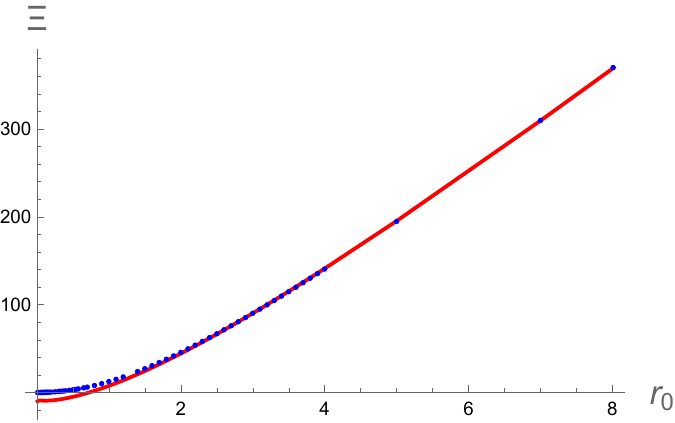}
    \caption{\small These plot show how the parameter $\Xi$ (dots) derived from the near-singularity geometry of the Kasner singularity depends on the horizon radius $r_0$. The left and right panels are associated with the Branch-1 and Branch-2 solutions respectively. The red solid lines describes the linear relations \eqref{xifitbranch1} and \eqref{xifitbranch2} respectively.}
    \label{XIplots}
\end{figure}

Finally, we would like to study the $\Xi$ parameter that can be read off at the Kasner singularity, via \eqref{xidef}. As we discussed above, the interior of the black holes of our Branch-1 solutions is close to the exact solution to Einstein-scalar theory of the scalar potential \eqref{scalarpot} with $\mu = -2\sqrt2/3$, for sufficiently large mass or $r_0$. For the exact solution we discussed in section \ref{sec:exact}, $\Xi$ was shown to be a linear function of the mass and the scalar charge, i.e.
\be
\Xi^{\rm analytic}= \ft{3}{23}(7 + 6\sqrt2)\, M - \fft{1}{23}(4\sqrt2-3)\,\Sigma\,.\label{xiMSig}
\ee
We find that this type of linear relation continues to hold, but with slightly different coefficients. Our data fitting suggests that $\Xi$, $M$ and $\Sigma$ are related by
\be
\Xi= 2.01944 M  -0.12783 \Sigma\,.\label{xifitbranch1}
\ee
In particular, for $r_0\ge 0.3$, the matching has errors less than $0.5\%$. In the left panel of Fig.~\ref{XIplots}, we plot the $\Xi$'s of Branch-1 solutions as a function of $r_0$. The dots represent the $\Xi$ read off directly from \eqref{xidef} and the solid line represents the $\Xi$ calculated from \eqref{xifitbranch1}. For $r_0\ge 0.3$, the accuracy of their match is within $0.5\%$.

\subsection{Branch-2 solutions}

\begin{figure}[ht]
    \centering
     \includegraphics[width=0.45\linewidth]{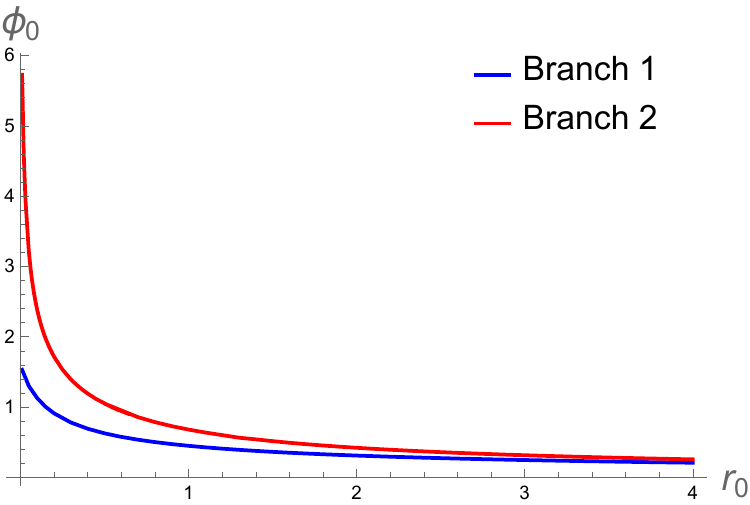}\ \ \
    \includegraphics[width=0.45\linewidth]{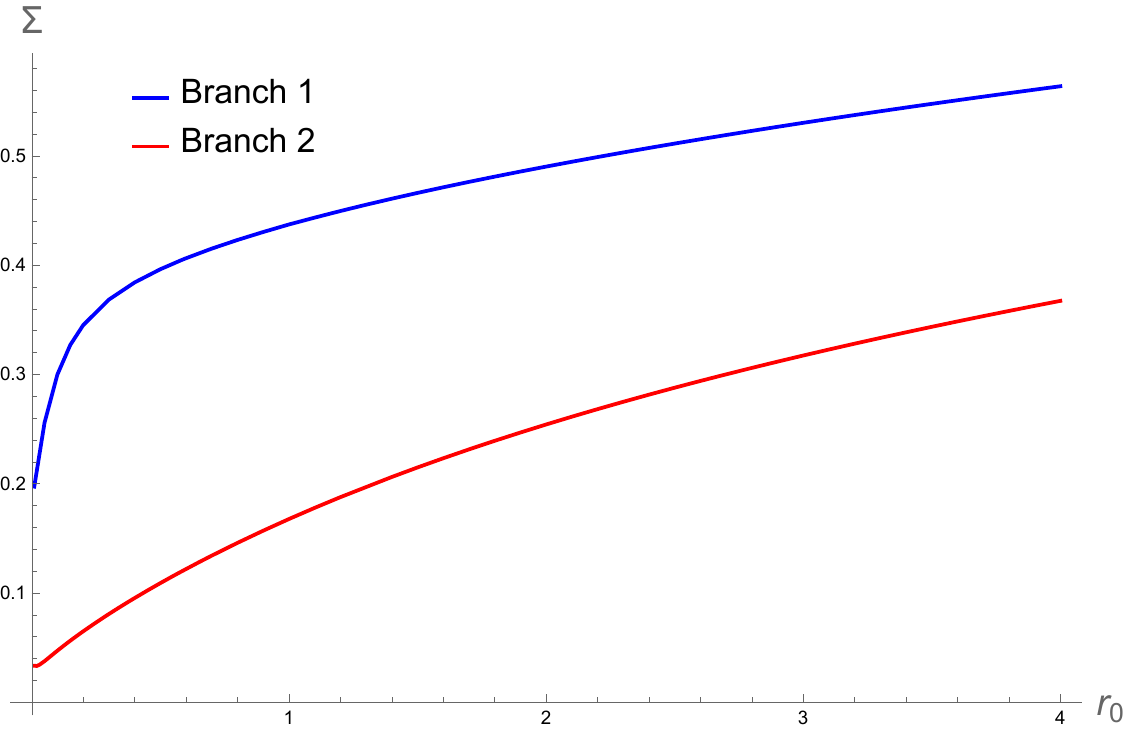}\ \ \
      \caption{\small The left and right panels give the horizon and asymptotic scalar hair as functions of the horizon radius. The two branches are sufficiently distinct in scalar hair, compared to the mass/radius relations, presented in Fig.~\ref{Mr02branch}.}
    \label{phi0Sig2branch}
\end{figure}

As we have mentioned earlier, for a given horizon radius $r_0$, there can be multiple black hole solutions. Here we shall not be repetitive, but present only the essential features of Branch-2 solutions. In Fig.~\ref{phi0Sig2branch}, we present both the horizon and asymptotic scalar hair parameters $(\phi_0, \Sigma)$ as functions of the horizon radius. In order to make a comparison to Branch-1 solutions, we present the plots for both branches. Interestingly, the size of $\phi_0$ and $\Sigma$ of the two branches reverses. The bigger the horizon hair $\phi_0$ yields a smaller asymptotic scalar charge $\Sigma$. The horizon scalar hair becomes degenerate and indistinguishable for the two branches at large $r_0$, but the asymptotic scalar charges remain distinct.

\begin{figure}[ht]
    \centering
     \includegraphics[width=0.45\linewidth]{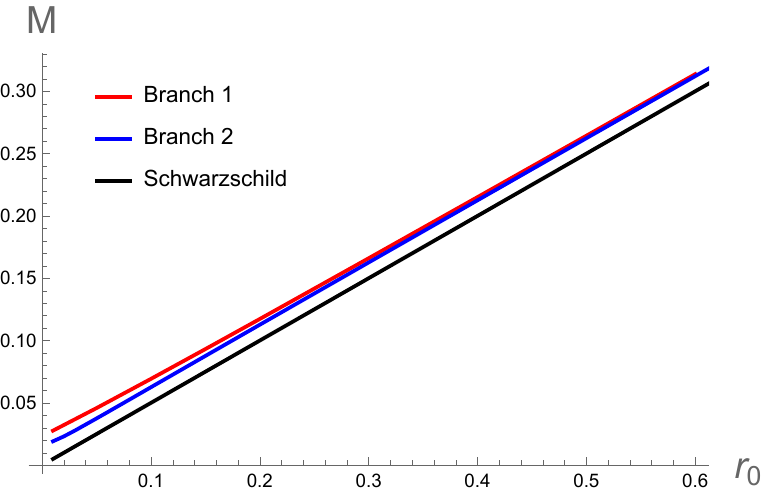}\ \ \
    \includegraphics[width=0.45\linewidth]{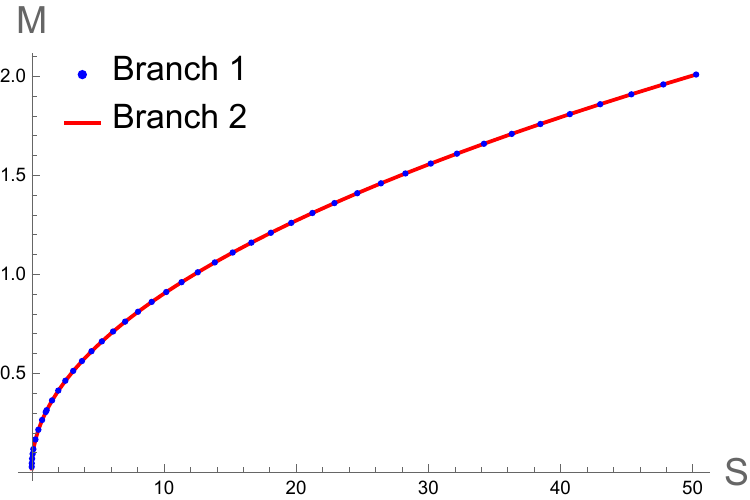}\ \ \
      \caption{\small Here are the mass/radius relations for both Branch-1 and Branch-2 solutions. They can be distinguished only for sufficiently small horizon radius, e.g.~$r_0<0.6$. In the right panel, we use dots for Branch-1 solutions and a solid line for Branch-2 solutions, so that one will not be totally covered by the other.}
    \label{Mr02branch}
\end{figure}

The mass/radius relation, on the other hand, is very degenerate for the two branches of solutions. In Fig.~\ref{Mr02branch}, we plot the relation for both branches.
Although they can be still decerned for sufficiently small radius, e.g.~$r_0\le 0.6$, the mass/radius relations become indistinguishable for larger $r_0$. We have to use dots for the Branch-1 solution and solid line for the Branch-2 solutions for distinctions in presentation. The degeneracy implies that their mass, entropy and temperature are also degenerate, so that for a given temperature, there can be an ensemble of different types of black holes of the similar thermodynamic properties.

Although different branches of black hole solutions have the degenerate mass/radius relation and hence they have the approximately the same exterior geometry, their interior structure are completely different from each other, as well as from the Schwarzschild black hole.
\begin{figure}[ht]
    \centering
     \includegraphics[width=0.3\linewidth]{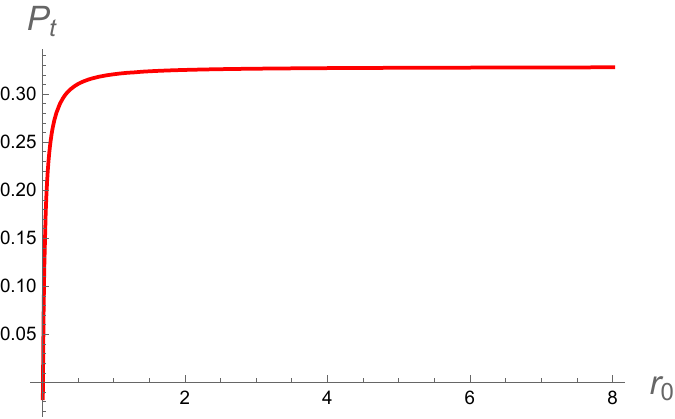}\ \ \
    \includegraphics[width=0.3\linewidth]{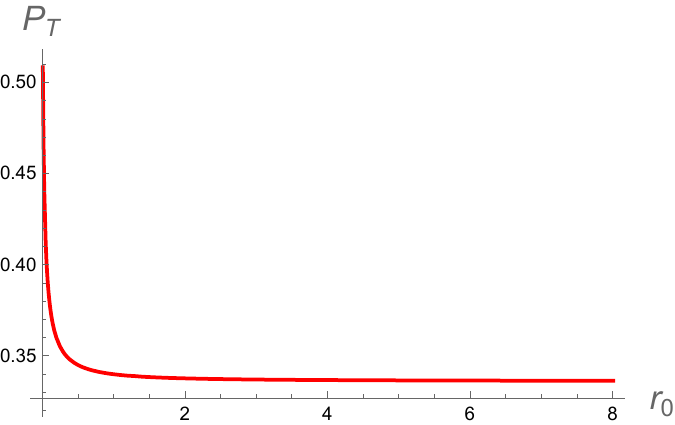}\ \ \
    \includegraphics[width=0.3\linewidth]{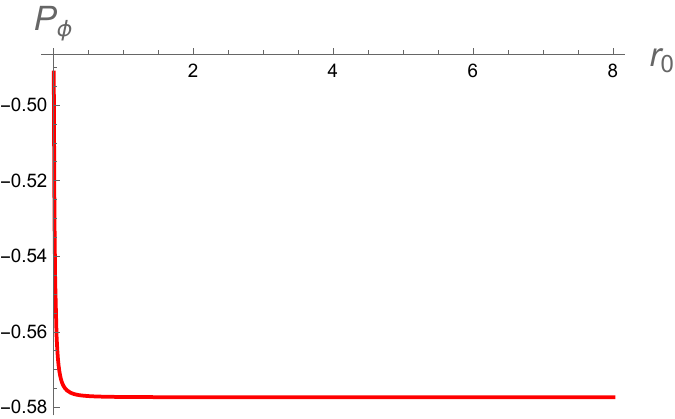}
    \caption{\small The Kasner exponents of Branch-2 solutions are functions of the black hole radius, but they approach asymptotically to constants at large $r_0$. These exponents are very different from those of Branch-1 solutions shown in Fig.~\ref{3Ps}, but they also satisfy the two identities in \eqref{Psconstraints} with extremely high accuracy ($\le 10^{-7}$.) Furthermore, $P_T$ is positive and bounded above as in \eqref{PTbound}.}
    \label{3Psbranch2}
\end{figure}
In Fig.~\ref{3Psbranch2}, we present the Kasner exponents $(P_t,P_T,P_\phi)$ as function of the horizon radius. They should be compared to those of the Branch-1 solutions, shown in Fig.~\ref{3Ps}, as well as to the Kasner exponents of the Schwarzschild black hole.

For the Branch-2 solutions, the linear dependence of $\Xi$ in terms of $M$ and $\Sigma$ does not hold for all masses. However, for sufficiently large $M$, the relation can still be accurate. We find, for $r_0\ge 2$, that the linear relation is
\be
\Xi=140.333 M - 383.199 \Sigma\,.\label{xifitbranch2}
\ee
This should be compared to \eqref{xifitbranch1}, which illustrate the stark difference of the two branches of solutions. The right-panel of Fig.~\ref{XIplots} illustrates explicitly that the linear relation indeed holds for $r_0\ge 2$.

Thus, we see that on the exterior, the Branch-1 and Branch-2 solutions are like twins with indistinguishable geometry, but deep inside, these solutions are starkly different.

\subsection{Miscellaneous branches}

\begin{figure}[ht]
    \centering
     \includegraphics[width=0.45\linewidth]{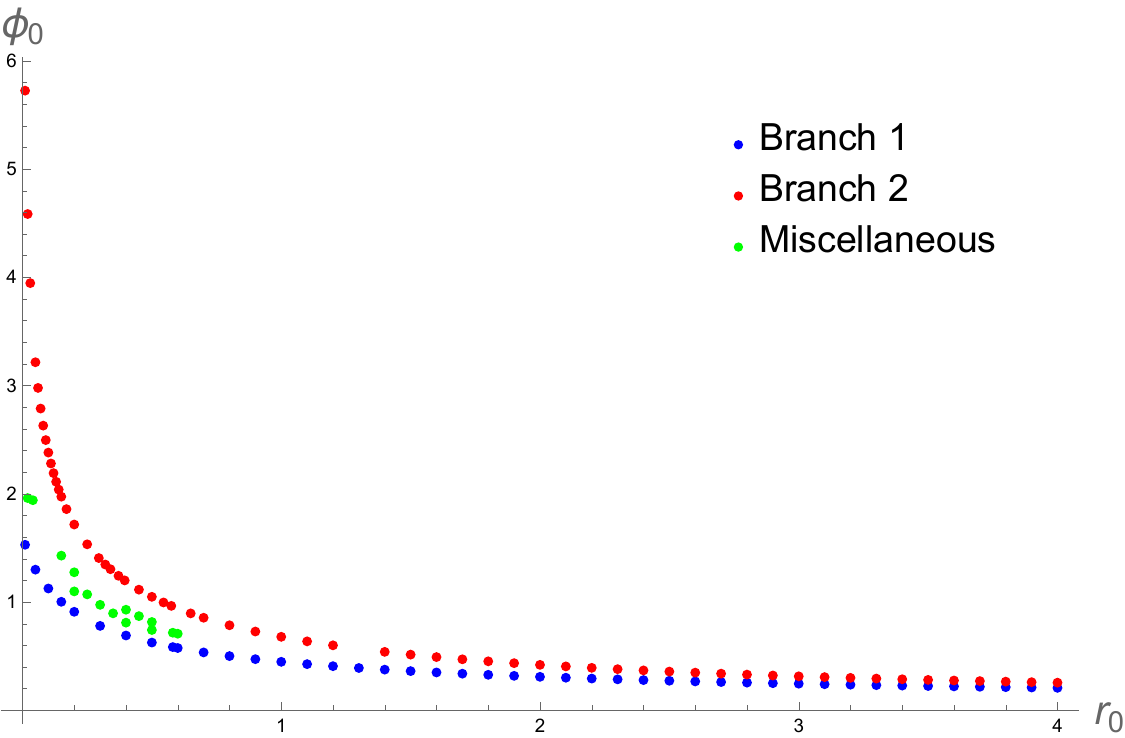}\ \ \
     \includegraphics[width=0.45\linewidth]{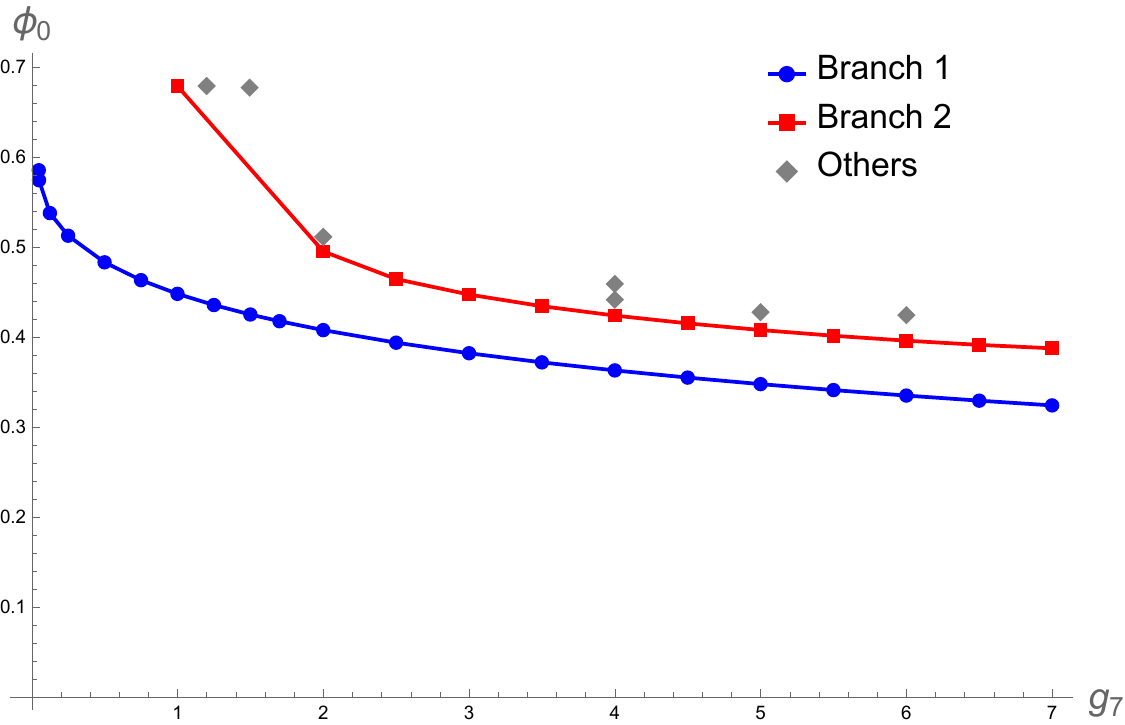}\ \ \
      \caption{\small The left panels shows additional miscellaneous solutions we find in the region where the gap between Branch-1 and Branch-2 solutions widens. In the right panel, we list scalar hairy black holes with fixed $g_5=1$, but vary $g_7$ for $r_0=1$. }
    \label{misc}
\end{figure}

We find further scalar hairy black holes for the scalar potential \eqref{g57} with fixed $g_5=1=g_7$. However, our numerical ability fails to give a coherent picture of these miscellaneous solutions. In the left panel of Fig.~\ref{misc}, we present the sporadic new black holes between the two branches.
As the $r_0$ approaches zero, the gap of horizon hair $\phi_0$ between the branches increase and new hairy black holes emerge. However, they emerge sporadically and we fail to find a general pattern. In the following tables (Tab.~\ref{tab1} and \ref{tab2}), we list some specific examples for $r_0=0.2$ and $r_0=0.5$ respectively. In both cases, we find a total of 5 black holes, including the Schwarzschild black hole, and Branch-1 and Branch-2 black holes discussed in detail.
\begin{table}
    \centering
    \begin{tabular}{cccccccc}
       \hline $\phi_0$\!\!\! & $M$ & $\Sigma$ & $T$ & $P_t$ & $P_T$ & $P_\phi$ & $\Xi$ \\
       \hline 1.71633\!\!\! & 0.112699  & 0.0647353 & 0.398192 & 0.280355 & 0.359823 & -0.575524 & 0.520871 \\
       \hline 1.27501\!\!\! & 0.117892 & 0.919848 & 0.386912 & 0.25968 & 0.37016 & -0.573816 & 0.541367\\
       \hline 1.09862\!\!\! & 0.117108 & 3.75577 & 0.385401 & 0.00617425 & 0.496913 & -0.503049 & 0.234276 \\
       \hline 0.910062\!\!\! & 0.117313 & 0.344872 & 0.385724  & -0.329979  & 0.66499  & -0.0578415 & 0.189607 \\
       \hline 0\!\!\! & 0.1 & 0 & 0.397887 & $-\ft13$ & $\ft23$ & 0 & 0.2 \\ \hline
    \end{tabular}
    \caption{Here is a list of discrete number of black holes at $r_0=0.2$. The bottom is the Schwarzschild black hole, the top and fourth belong to Branch-2 and Branch-1 solutions. In between are the two miscellaneous solutions.}
    \label{tab1}
\end{table}
\begin{table}[ht]
\centering
    \centering
    \begin{tabular}{cccccccc}
       \hline $\phi_0$ & $M$ & $\Sigma$ & $T$ & $P_t$ & $P_T$ & $P_\phi$ & $\Xi$ \\
       \hline 1.04845& 0.262338& 0.108855& 0.158515& 0.310243& 0.344879& -0.577004&  3.10598 \\
       \hline 0.816041& 0.26513& 2.07915& 0.157195& 0.295623& 0.352188& -0.576426& 2.76971\\
       \hline 0.741882& 0.264241& 3.52222& 0.15703& 0.242077& 0.378962& -0.571916& 1.53418  \\
       \hline 0.626022  & 0.264343 & 0.396089 & 0.157021  & -0.326556  & 0.663278  & -0.0821154 & 0.483709 \\
       \hline 0 & 0.25 & 0 & 0.159155 & $-\ft13$ & $\ft23$ & 0 & 0.5 \\
       \hline
    \end{tabular}
    \caption{Here is a list of discrete number of black holes with $r_0=0.5$. The bottom is the Schwarzschild black hole, the top and fourth belong to Branch-2 and Branch-1 solutions. In between are the two miscellaneous solutions.}
    \label{tab2}
\end{table}

\subsection{Varying coupling constants}

We have so far reported numerical data for scalar hairy black holes associated with the scalar potential \eqref{g57} with fixed $g_5=1=g_7$. We have also constructed black holes for varying $g_7$, but with fixed $g_5=1$ and $r_0=1$. Since it is a time consuming and involved business, we do not intend to study them systematically here, but only illustrate their existence by presenting the data in the right-panel of Fig.~\ref{misc}. Even for fixed $g_5$ and $r_0$, we can see that a rich spectrum emerge for such a simple scalar potential \eqref{g57}. A more thorough investigation with more advanced numerical technique is called for.

\section{Interior geodesics and maximum surviving time}
\label{sec:interiorgeo}

We have so far studied a variety of scalar hairy black holes in Einstein-scalar gravity. We studied their interior geometry focused on Kasner singularity, characterized by the Kasner exponents and the parameter $\Xi$ read off from the near-singularity geometry for the scalar kinetic-term dominated solutions. We now consider the geodesic probes of the interior cosmological geometry. Once a particle enters the event horizon of a black hole with a spacelike singularity, it will inevitably fall into the singularity. One important parameter characterizing the black hole interior is the longest proper time that a massive particle can survive inside a black hole. For the spherically-symmetric and static black holes of the type \eqref{newmetric}, it is given by
\be
\tau = \int_0^{r_0} \fft{dr}{\sqrt{-f}}\,.
\ee
For the Schwarzschild black hole of mass $M$, it is simply given by $\tau_{\rm sch} = \pi M$. We can thus calculate the dimensionless maximum surviving time $\tau/(\pi M)$ inside a black hole of mass $M$ as one characterization of its interior geometry.

\begin{figure}[ht]
    \centering
     \includegraphics[width=0.45\linewidth]{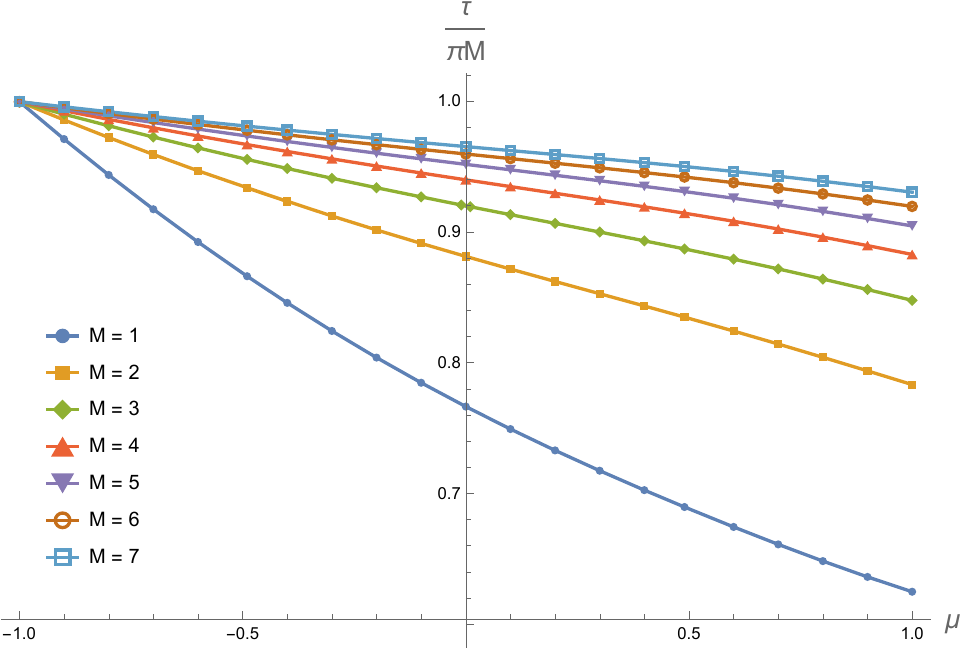}\ \ \
     \includegraphics[width=0.45\linewidth]{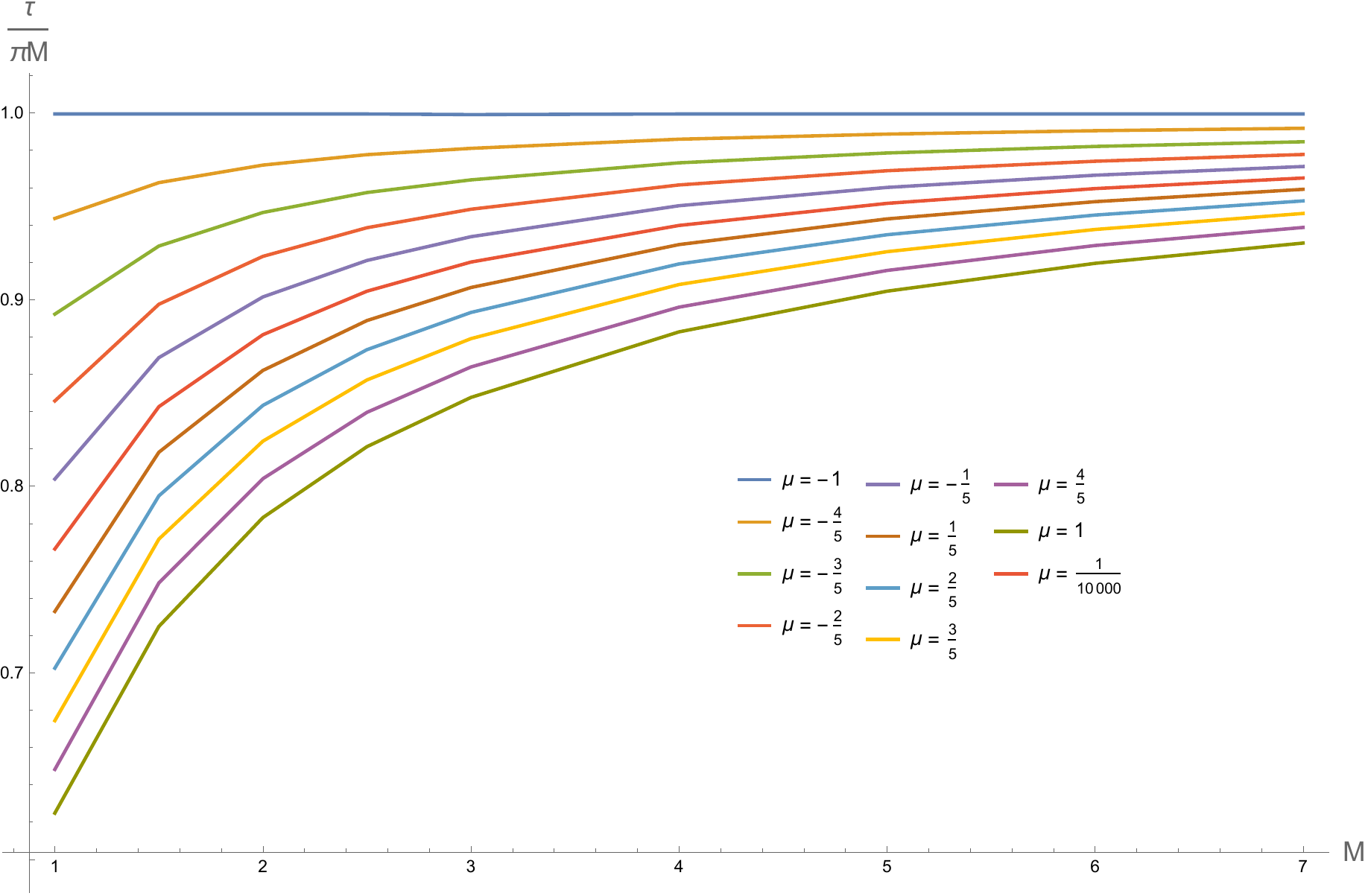}
    \caption{\small Here we give the maximum dimensionless surviving time of a massless particle inside the exact scalar hairy black holes of Section \ref{sec:exact}. For the Schwarzschild black hole, this quantity is 1. All the black holes satisfy the inequality \eqref{inequality}. }
    \label{ExactSolTaumu}
\end{figure}

We first consider the exact scalar hairy black holes discussed in section \ref{sec:exact}. In Fig.~\ref{ExactSolTaumu}, we plot how the dimensionless time $\tau/(\pi M)$ depend on the mass and also the theory parameter $\mu$.
We see that in all the cases, we have an inequality $\tau/(\pi M)\le 1$, with the Schwarzschild black hole saturating the bound. Furthermore, for all different $\mu$ values, the dimensionless time $\tau/(\pi M)$ approaches 1 as the mass increases. These data indicate the following interior structures of scalar hairy black holes. On one hand, as we have seen earlier, as mass increases, the exterior geometry of the black hole starts to resemble the Schwarzschild black hole as the mass becomes large; however, the Kasner singularity remains resolutely different, independent of the mass. On the other hand, as the mass increases, the fact that $\tau/(\pi M)$ approaches 1 as mass become large indicates that the maximum surviving time of a particle in the interior for a large black hole is mostly spent near the horizon whose geometry resembles that of the Schwarzschild black hole. The time portion of the ``final life'' spent near the Kasner singularity becomes insignificant for the large-mass black hole.

\begin{figure}[ht]
    \centering
     \includegraphics[width=0.45\linewidth]{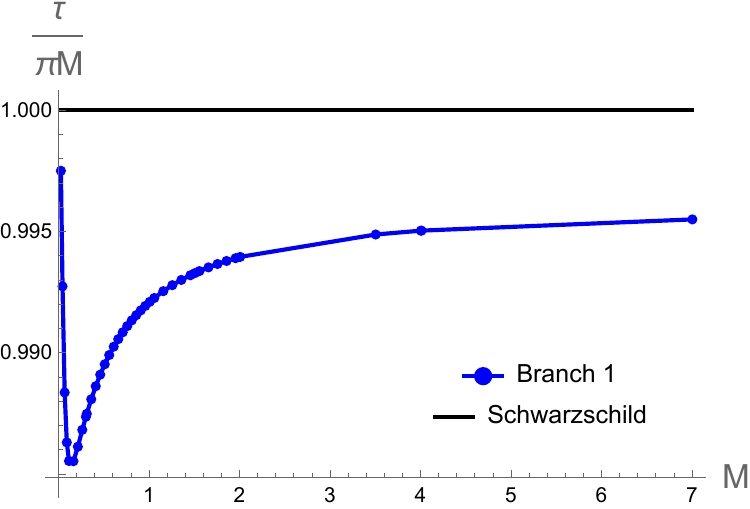}\ \ \
     \includegraphics[width=0.45\linewidth]{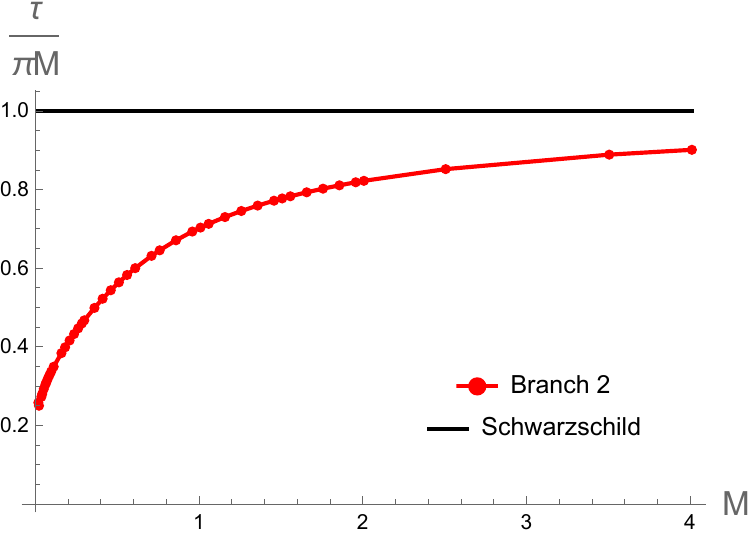}
    \caption{\small Here we plot the maximum dimensionless surviving time of a massless particle inside the branch-1 and branch-2 black holes. The branch-1 solution resembles more the Schwarzschild black hole. All the examples satisfy the inequality \eqref{inequality}.}
    \label{numericalSolTaumu}
\end{figure}

In Fig.~\ref{numericalSolTaumu}, we present the dimensionless maximum surviving time
$\tau/(\pi M)$ for the Branch-1 and Branch-2 solutions of the numerical constructions in Section \ref{sec:numerical}.
The Branch-1 solution resembles the Schwarzschild black hole much more than the Branch-2 solutions. Nevertheless, we see that for both branches, we have again $\tau/(\pi M)\le 1$.

\section{Conclusions}
\label{sec:conclusion}

In this paper, we considered Einstein gravity minimally coupled to a real massless scalar with a suitable scalar potential that admits spherically-symmetric and static scalar hairy black holes that are asymptotic to the Minkowski spacetime in four dimensions. We studied the interior structure of these black holes, focusing on the properties and classifications of the their Kasner singularity.

We avoided considering asymptotic AdS geometries since the falloff structures of both the scalar and metric functions are less restrictive in the AdS cases. For the asymptotic Minkowski spacetime, the leading falloffs for a general solution involves two independent parameters, the mass $M$ and the scalar hair $\Sigma$, defined in \eqref{twop}. The weak version of the black hole no-hair theorem does not rule out the scalar field, but restricts it so that the scalar hair $\Sigma$ is not an independent parameter, but a function of the mass, i.e.~$\Sigma=\Sigma(M)$. For a system that does not admit exact solutions, or exact solutions are unknown, determining the proper black hole $\Sigma(M)$ is a tedious fine-tuning process that one must go through to establish before we can study their internal structures.

In this paper, we began with an analysis of a class of known exact black hole solutions in literature. After examining their thermodynamic properties based on the exterior structures, we analysed their interior Kasner singularity. This class of exact solutions are particularly inclusive, with both the scalar kinetic dominating solutions at the singularity and those that are not. In either cases, we found that the three Kasner exponents $\{P_t,P_T,P_\phi\}$ all depend only on the theory parameter, but independent of the mass of the black hole. As one would have expected, for the kinetic dominating solutions, the Kasner exponents satisfy the constraints \eqref{Psconstraints}, while for the non-kinetic dominating solutions, there are no such constraints. An intriguing feature from the exact solutions is that there is a black hole mass gap for kinetic dominating black holes while for non-kinetic dominating black holes, such a mass gap does not exist. It is certainly of great interest to investigate whether this is a universal feature, and if so, why the properties in the singularity dictates whether the black hole has a mass gap or not. We also observed that $P_T$ is always positive and bounded above, as in \eqref{PTbound}.

In addition to the Kasner exponents, the asymptotic information $(M,\Sigma(M))$ can also be extracted from the near-singularity geometry. Independent of whether the singularity is kinetic dominating or not, we could extract a dimensionless constant $\Theta$ from the metric function at the singularity, as defined in \eqref{Thetadef}. For the Schwarzschild black hole, this quantity is simply 1. For the scalar hairy black holes, we found that this parameter is a function of the dimensionless ratio $M/\Sigma$. We obtained the exact expression $\Theta(M/\Sigma)$ for the exact solutions in section \ref{sec:exact}. Unfortunately, these explicit expressions are not enlightening. A more intriguing constant, extract from the kinetic dominating singularity, is $\Xi$ defined in \eqref{xidef}. From the exact solutions, we found that the constant $\Xi$ is a linear function of $M$ and $\Sigma$.

As exact solutions could be too special and not all representative, we then considered more general scalar potentials where there are no exact black hole solutions. We considered the scalar potential \eqref{g57} and construct new numerical black hole solutions. Despite the simplicity of \eqref{g57}, the Einstein-scalar gravity admits a richer spectrum of scalar hairy black hole solutions. For a given horizon radius (and hence entropy), we found that multiple scalar hairy black holes exist. In other words, there is a discrete set of scalar hair parameter $\Sigma(M)$. In contrast, we found no numerical evidence for multiple hairy black holes with given mass in Einstein-scalar gravity with potential \eqref{scalarpot}, where exact solutions can be constructed.

Unlike the black holes with exact solutions, in the more general numerical constructions, the Kasner exponents themselves can depend on the mass of the black hole; however, they asymptote some constant values as mass increases. We found that
\be
\lim_{M\rightarrow \infty} \{P_t,P_T,P_\phi\} = \{P_t^*, P_T^*, P_\phi^*\}\,.
\ee
Furthermore, our numerical data suggest the following large-mass asymptotic behavior
\be
\lim_{M\rightarrow \infty} \Xi = \xi_1 M + \xi_2 \Sigma\,,\qquad
\hbox{with}\qquad \xi_1 \ge 2\,.
\ee
The inequality is saturated by the Schwarzschild black hole. We therefore tentatively answered the two questions raised in the introduction, and expect that these may be generally true for scalar hairy black holes in Einstein-scalar theories. An important lesson from this result is that although scalar hairy black holes can be geometrically similar to the Schwarzschild black hole (or other scalar hairy black holes) in the exterior, its interior structure, the near-singularity geometry in particular, can be completely different.

Finally, we probed the interior geometry of the scalar hairy black holes by studying the geodesic motion of a massive particle. We found that, based on surveying black hole examples, once the particle entered the horizon, the maximum surviving time before it fell into the Kasner singularity satisfied the inequality
\be
\tau \le \pi M\,.\label{inequality}
\ee
The Schwarzschild black hole saturates the bound. Our work illustrates that even for simple Einstein-scalar gravity, the interior of the scalar hairy black holes have rich structures that deserve a more thorough investigation.

\section*{Acknowledgement}

This work is supported in part by the National Natural Science Foundation of China (NSFC) grants No.~12375052 and No.~11935009, and also by the Tianjin University Self-Innovation Fund Extreme Basic Research Project Grant No.~2025XJ21-0007.

\appendix

\section{Kasner singularity from higher dimensions}
\label{sec:app}

In this paper, we focused on the scalar hairy black holes with Kasner singularity where the scalar kinetic energy dominate over the potential energy. The relevant Lagrangian in the singularity region is
\be
{\cal L}_4= \sqrt{-g} \Big(R -\ft12 (\partial\phi)^2\Big).\label{d4lag}
\ee
This four-dimensional Lagrangian can be obtained from Einstein gravity $\hat {\cal L}=\sqrt{-\hat g} \hat R$ in a generic $D=4+n$ dimensional spacetime, reduced on an $n$-torus, keeping only the breathing scalar mode, namely \cite{Bremer:1998zp}
\be
d\hat s_D^2 =  e^{2\alpha\phi} ds_4^2 + e^{2\beta\phi} ds_{T^n}^2\,,\qquad
\alpha^2 = \fft{n}{4(n+2)}\,,\qquad \beta = -\fft{2\alpha}{n}\,.\label{reduction}
\ee
The relevant Kasner spacetime that fits the above reduction ansatz takes the form
\be
d\hat s_D^2 = - d\hat \tau^2 + a_1^2 \hat \tau^{{2\hat P}_t} dt^2 + a_2^2 \hat \tau^{2 {\hat P}_T} d\Omega_2^2 + \hat\tau^{2{\hat P}_z} (dz_1^2 + \cdots + dz_n^2)\,.\label{highDsol}
\ee
The Ricci flatness at $\hat \tau\rightarrow 0$ requires that the Kasner exponents must satisfy
\be
{\hat P}_t + 2{\hat P}_T + n {\hat P}_z =1\,,\qquad
{\hat P}_t^2 +2 {\hat P}_T^2 + n {\hat P}_z^2=1\,.\label{highDcons}
\ee
This is a special case of the most general anisotropic Kasner universe in higher dimensional pure gravity
\be
ds^2 = -d\hat \tau^2 + \sum \hat \tau^{2\hat P_i} dz_i^2\,,\qquad
\sum_i {\hat P}_i=1\,,\qquad \sum_i {\hat P}_i^2 =1\,.
\ee
Comparing the four and higher dimensional Kasner singularities \eqref{D4kasnersol} and \eqref{highDsol}, as well as the reduction ansatz \eqref{reduction}, we have
\be
\hat P_t = \fft{\sqrt{n+2} P_t - \sqrt{n}P_\phi}{\sqrt{n+2}-\sqrt{n} P_\phi}\,,\qquad
\hat P_T =  \fft{\sqrt{n+2} P_T - \sqrt{n}P_\phi}{\sqrt{n+2}-\sqrt{n} P_\phi}\,,\qquad
\hat P_z = \fft{2P_\phi}{\sqrt{n(n+2)} - n P_\phi}\,.
\ee
It is then easy to verify that the higher-dimensional constraints \eqref{highDcons} on the
Kasner exponents lead to the four-dimensional constraints in \eqref{Psconstraints}. As we remarked in Section \ref{sec:ES}, Kasner-type solutions \eqref{D4kasnersol} and \eqref{highDsol} are only the leading-order approximate solution in the $\tau\rightarrow 0$ limit to \eqref{d4lag} and pure gravity respectively; they becomes exact solutions for all $\tau$ if we replace the 2-sphere metric $d\Omega_2^2$ by a Ricci-flat metric.

\end{document}